%% file: main.tex
\documentclass[%
  reprint,
  amsmath,
  amssymb,
  aip,
]{revtex4-2}

\usepackage{listings}
\usepackage{graphicx}
\usepackage{dcolumn}
\usepackage{bm}
\usepackage{mathtools}
\usepackage{multirow}
\usepackage{siunitx} 
\DeclareSIUnit[number-unit-product = {\,}]
\cal{cal}

\usepackage[hidelinks]{hyperref}

\usepackage[normalem]{ulem} 

\usepackage{tikz}

\newcommand{\subref}[2]{\hyperref[#1]{\ref*{#1}#2}} 
\newcommand{\pness}{p^\mathrm{ss}}

\begin{document}

\title{
  Data-Efficient Multidimensional Free Energy Estimation via Physics-Informed Score Learning
}

\author{Daniel Nagel}
\email{nagel.phys@gmail.com}
\affiliation{%
  Institute for Theoretical Physics, Heidelberg University, 69120 Heidelberg, Germany
}%
\author{Tristan Bereau}%
\affiliation{%
  Institute for Theoretical Physics, Heidelberg University, 69120 Heidelberg, Germany
}%
\affiliation{%
  Interdisciplinary Center for Scientific Computing (IWR), Heidelberg University, 69120 Heidelberg, Germany
}%

\date{\today}

\begin{abstract}
  Many biological processes involve numerous coupled degrees of freedom, yet
  free-energy estimation is often restricted to one-dimensional profiles to
  mitigate the high computational cost of multidimensional sampling.
  In this work, we extend Fokker--Planck Score Learning (FPSL) to efficiently reconstruct
  two-dimensional free-energy landscapes from non-equilibrium molecular dynamics
  simulations using different types of collective variables.
  We show that explicitly modeling orthogonal degrees of freedom reveals insights
  hidden in one-dimensional projections at negligible computational overhead.
  Additionally, exploiting symmetries in the underlying landscape enhances
  reconstruction accuracy, while regularization techniques ensure numerical
  robustness in sparsely sampled regions. We validate our approach on three
  distinct systems: the conformational dynamics of alanine dipeptide, as well as
  coarse-grained and all-atom models of solute permeation through lipid
  bilayers. We demonstrate that, because FPSL learns a smooth score function
  rather than histogram-based densities, it overcomes the exponential scaling of
  grid-based methods, establishing it as a data-efficient and scalable tool for
  multidimensional free-energy estimation.
\end{abstract}

\maketitle

\input{body}

\section*{Acknowledgments}
We dedicate this work to Kurt Kremer on the occasion of his 70th birthday---an
inspiring and cherished mentor to T.B. The community owes a great debt to
Kremer's seminal work in soft-matter physics and multiscale simulations,
systematically aiming for simple, clear, and elegant modeling choices to further
our understanding of complex liquids.

We are grateful to Jeffrey Comer for helpful discussions
and Luis J. Walter and Sander Hummerich for their valuable feedback on the
manuscript.

We acknowledge support by the Deutsche Forschungsgemeinschaft (DFG, German
Research Foundation) under Germany's Excellence Strategy EXC 2181/1 - 390900948
(the Heidelberg STRUCTURES Excellence Cluster) and Heidelberg University through
the Research Council of the Field of Focus 2 ``Patterns and Structures in
Mathematics, Data, and the Material World.''

\section*{Data availability statement} A Python package implementing the
Fokker--Planck Score Learning framework is available under an open-source
license at
\url{https://github.com/BereauLab/fokker-planck-score-learning}.

\bibliography{lit_iso4}

\end{document}

%% file: body.tex
\section{Introduction}

Free-energy landscapes provide the thermodynamic map for molecular processes,
describing the stability of states and the kinetics of transitions between
them.\cite{chipot2007free, hansen2014practical, frenkel2023understanding} While
these landscapes are inherently high-dimensional, defining them is typically
done by projecting the complex configurational space onto a small set of
collective variables (CVs). In practice, this projection is often restricted to
a single dimension to minimize computational cost. However, this reduction
relies on the strong assumption that all orthogonal degrees of freedom
equilibrate rapidly. When this timescale separation fails, one-dimensional profiles
suffer from hysteresis, hidden barriers, and systematic errors.\cite{best2005reaction}
Extending the analysis to two dimensions is often necessary to resolve these
artifacts, yet robustly estimating two-dimensional (2D) free-energy surfaces remains a significant
computational bottleneck.

Equilibrium sampling approaches remain foundational for free-energy
calculations. Metadynamics adaptively deposits history-dependent bias to escape
deep minima,\cite{laio2002escaping} adaptive biasing force (ABF) methods apply
on-the-fly forces to flatten barriers,\cite{darve2008adaptive} while other
methods, such as replica-exchange\cite{sugita1999replica} and accelerated
molecular dynamics,\cite{hamelberg2004accelerated} can further enhance sampling
efficiency. Nevertheless, umbrella sampling--based methods remain the current
gold standard: by applying overlapping harmonic restraints, umbrella
sampling\cite{torrie1977nonphysical, roux1995calculation} ensures comprehensive
coverage of high-energy regions, and the weighted histogram analysis method
(WHAM)\cite{kumar92, hub10} or multistate Bennett acceptance ratio
(MBAR)\cite{shirts2008} estimators reconstruct the unbiased potential of mean
force (PMF) from the biased samples. These methods have become a workhorse for
various biomolecular applications, ranging from protein-ligand
binding\cite{orsi2010passive, buch2011optimized, swift2013back} to solute
permeation through lipid bilayers.\cite{carpenter2014method, lee2016simulation,
bennion2017predicting, tse2018link, menichetti2017silico, menichetti2019drug}
However, these conventional methods face a steep trade-off between accuracy and
scaling. Grid-based approaches like umbrella sampling suffer from the ``curse of
dimensionality'' starting from 2D, where the number of overlapping windows required to
cover the grid grows exponentially. Conversely, non-equilibrium methods based on
Jarzynski's equality\cite{jarzynski1997nonequilibrium} theoretically apply to
any dimension but struggle to converge when work distributions are broad,
limiting their practical utility for complex landscapes.

\begin{figure}[t]
  \centering
  \includegraphics[width=\linewidth]{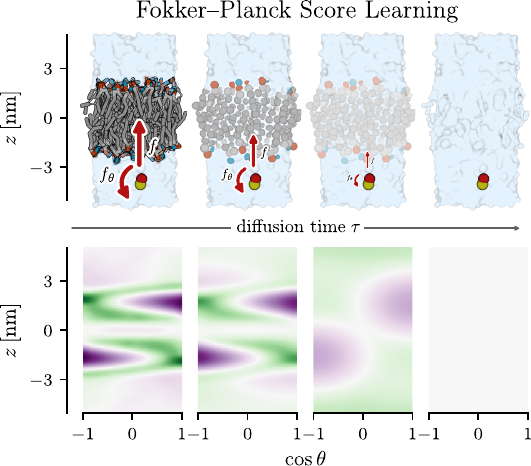}
  \caption{
    We consider a system with periodic boundary conditions, characterized by a
    conservative potential of mean force, $U(z, \theta)$. The center of mass of
    the composite particle (red and yellow) is driven by a constant external
    force $f$, while its orientation is subject to a constant torque $f_\theta$.
    The steady-state solution of the
    Fokker--Planck equation for a Brownian particle in a periodic potential,
    $\pness$, informs the score of our diffusion model, mapping the
    non-equilibrium steady state to the equilibrium distribution. The diffusion
    model smoothly interpolates between the physical non-equilibrium system at
    constant flux $J$ (left) and a trivial uniform prior (right). Denoising
    allows us to efficiently reconstruct the two-dimensional equilibrium
    free-energy landscape (bottom row) by exploiting the structure of $\pness$.
  }
  \label{fig:method}
\end{figure}

In this work, we address these limitations by extending the recently introduced \emph{Fokker--Planck Score Learning} (FPSL)\cite{nagel2025fokker} to multidimensional free-energy
estimation. FPSL frames free-energy reconstruction as a generative modeling
task, training a diffusion model to learn the equilibrium free-energy landscape
from non-equilibrium trajectory data. Crucially,
FPSL distinguishes itself by explicitly exploiting the periodicity of the
collective variables: it embeds the analytic non-equilibrium steady state (NESS)
of a periodic driven system directly into the training objective.
This physics-informed prior acts as a powerful inductive bias,
allowing one to learn equilibrium free-energy landscapes from non-equilibrium data with high
efficiency. For an illustration of the method, see Fig.~\ref{fig:method}.

Because FPSL is a score-based method, it generalizes readily to higher
dimensions without structural changes or significant computational overhead in
the learning phase. The neural network simply learns a smooth vector field in a
higher-dimensional space. The only added complexity lies in the physical
sampling: higher-dimensional spaces naturally require longer molecular dynamics
(MD) simulations to sufficiently sample the NESS distribution.
However, we show that this investment is highly efficient. We demonstrate that
even when the primary interest lies in a one-dimensional profile, it is
advantageous to learn the full 2D landscape and subsequently marginalize over
the orthogonal degree of freedom. This approach resolves hidden barriers and
yields faster convergence than direct 1D estimation or conventional umbrella
sampling-based MBAR.

We validate the versatility of this framework by applying it to three distinct
systems: the conformational dynamics of alanine dipeptide in water,
described by two dihedral angles; the permeation of a solute
through a lipid bilayer modeled with a coarse-grained force field, which
introduces a non-periodic orientational degree of freedom alongside a periodic
spatial coordinate; and the permeation of ethanol through a membrane modeled
in all-atom detail.
These applications highlight the generality of FPSL: the method is agnostic to
the nature of the collective variables---whether Cartesian coordinates such as
the position within the simulation box or internal coordinates such as dihedral
angles---and to the force field resolution. Because FPSL relies solely on the
ability to efficiently sample the stationary probability density in the CV
space, which is naturally facilitated by the periodicity of the driving
coordinate, it requires no system-specific adaptations. This provides a unified
and efficient route to multidimensional free-energy estimation
of collective variables that can be mapped to periodic domains.

The remainder of this paper is organized as follows. Sec.~\ref{sec:theory}
recalls the derivation of Fokker--Planck Score Learning and details its extension to 2D
systems, including the enforcement of symmetries and the use of Fokker--Planck
regularization.
Sec.~\ref{sec:results} presents the results for the three test cases, demonstrating
the superior convergence and accuracy of our 2D approach compared to standard baselines.
Finally, we summarize our findings in Sec.~\ref{sec:conclusion}.

\section{Theory and Methods\label{sec:theory}}

\subsection{Background\label{sec:background}}

We briefly review the theoretical foundations of score-based diffusion models
and the non-equilibrium steady state (NESS) for periodic systems, which
together form the basis of Fokker--Planck Score Learning
(FPSL).\cite{nagel2025fokker} For comprehensive derivations, we refer to
Song et al.\cite{song2020score} and Risken.\cite{risken1996fokker}

\subparagraph{Score-Based Diffusion Models on Periodic Domains:}
Denoising diffusion models (DDMs) reconstruct a target distribution
$p_{\text{data}}(x)$ by inverting a gradual noising
process.\cite{sohldickstein15,ho20,song2020score} On a periodic domain of
size $L$, the forward process is a Brownian motion on the torus,
\begin{align}
  \mathrm{d}x_\tau &= \sqrt{2\alpha_\tau}\mathrm{d}W_\tau\,, \label{eq:ddm_fwd_sde_periodic}
\end{align}
where $\tau \in [0, 1]$ is the dimensionless diffusion time and $\alpha_\tau >
0$ the noise schedule. As $\tau \to 1$, the distribution relaxes to a
uniform prior over $[-L/2, L/2)$. The corresponding reverse-time
SDE\cite{anderson82}
\begin{align}
  \mathrm{d}x_\tau &= -2\alpha_\tau\nabla \ln p_\tau(x_\tau)\mathrm{d}\tau + \sqrt{2\alpha_\tau}\mathrm{d}\bar{W}_\tau\,, \label{eq:ddm_rev_sde_periodic}
\end{align}
maps the uniform prior back to the data distribution, with the score function
$s(x_\tau,\tau) = \nabla \ln p_\tau(x_\tau)$ governing the drift.

The score is learned via denoising score matching.\cite{song2020score}
Noised samples $x_\tau = x_0 + \sigma_\tau \varepsilon$, with
$\varepsilon \sim \mathcal{N}(0,1)$ and $\tau\sim \mathcal{U}([0,1])$, are wrapped
onto the periodic domain, and
a neural network $\varepsilon_\theta(x_\tau, \tau)$ is trained to predict the
added noise by minimizing
\begin{equation}
  \mathcal{L}_\text{DSM}(\theta) = \mathbb{E}_{x_0\sim p_0, \varepsilon, \tau} \left[ \left\| \varepsilon_\theta(x_\tau, \tau) - \varepsilon \right\|^2 \right]\,,
  \label{eq:dsm_loss}
\end{equation}
where the relationship $s_\theta(x_\tau, \tau) = -\varepsilon_\theta(x_\tau,
\tau) / \sigma_\tau$ recovers the score from the predicted noise.
To ensure that the learned score corresponds to the gradient of a scalar
potential, we employ energy-based diffusion
models,\cite{arts2023two, mate24, mate2025solvation} parameterizing the noise
as
\begin{equation}
  \frac{\varepsilon_\theta(x_\tau, \tau)}{\sigma_\tau}
  = - s_\theta(x_\tau, \tau)
  = \beta \nabla U_\theta(x_\tau, \tau)\,,
  \label{eq:energy_score}
\end{equation}
where $U_\theta$ is a learnable, time-dependent potential energy function
(in units of $k_\mathrm{B}T$) and $\beta = (k_\mathrm{B}T)^{-1}$. The
gradient is computed via automatic differentiation.

\subparagraph{Non-Equilibrium Steady State of a Periodic System:}
Consider a particle undergoing overdamped Langevin dynamics in a periodic
potential $U(x)$ subject to a constant external force $f$. Defining the
effective (tilted) potential $U_\text{eff}(x) = U(x) - fx$, the corresponding
Fokker--Planck equation admits a time-independent solution with constant
probability current.\cite{risken1996fokker} Specifically, the probability
density $p_t(x)$ evolves according to
\begin{equation}\label{eq:fokker_planck}
  \begin{aligned}
    \partial_t p_t(x) &= \nabla\left\{ \beta D \nabla U_\text{eff}(x)\, p_t(x)\right\} + D\,\nabla^2 p_t(x)\\
    &= -\nabla J_t(x)\,,
  \end{aligned}
\end{equation}
where $D$ is the diffusion coefficient and $J_t(x)$ is the probability current.
In the steady-state limit, the current becomes spatially uniform, $J_t(x) = J =
\text{const}$.

For a system with period $L$, the NESS distribution reads\cite{nagel2025fokker}
\begin{equation}
  \pness(x) \propto e^{-\beta U_\text{eff}(x)}\, \int_{x}^{x+L} \textrm{d}y\: e^{\beta U_\text{eff}(y)}\,.\label{eq:ness_pbc}
\end{equation}
The first factor is the local Boltzmann weight of the tilted potential, while
the integral enforces the periodic boundary conditions and guarantees a constant
probability current. Without periodicity, the distribution reduces to the
tilted Boltzmann form $\pness(x) \propto e^{-\beta U_\text{eff}(x)}$.

\subsection{Fokker--Planck Score Learning\label{sec:fpsl}}
To embed the NESS physics into the diffusion model, we define a
diffusion-time-dependent effective potential that interpolates between the
physical system and the uniform prior over $\tau \in [0, 1]$
\begin{equation}\label{eq:U_eff_dm}
  U_\text{eff}(x, \tau) = (1-\tau) \left[U_\tau(x) - fx\right] + \tau C_\text{prior}\,,
\end{equation}
where $U_\tau(x)$ is the potential energy at diffusion time $\tau$,
$f$ is the constant external driving force, and $C_\text{prior}$ is a
normalization constant. All potentials throughout this work are expressed in
units of $k_\mathrm{B}T$. At $\tau = 0$, this reduces to the physical
effective potential, while at $\tau = 1$ it yields the uniform prior distribution
on the periodic domain, characterized by the constant $C_\text{prior}$.

The objective is to train a neural network $U^\theta_\tau(x)$ such that at the
initial diffusion time ($\tau=0$), $U^\theta_0(x)$ approximates the physical
potential of mean force $U(x)$. Leveraging the NESS distribution,
Eq.~\eqref{eq:ness_pbc}, and the energy-based score parameterization, the score
function takes the form
\begin{align}
  s^\theta(x_\tau, \tau) =&\nabla \ln \pness(x_\tau, \tau)\nonumber\\
  =&-\beta \nabla U^\theta_\text{eff}(x_\tau, \tau) + \Delta s^\theta(x_\tau, \tau)\label{eq:score_with_correction}\,,
\end{align}
where $U^\theta_\text{eff}(x_\tau, \tau)$ is the neural network approximation of the
effective potential, while only the equilibrium potential itself is parametrized
by the network, i.e., $U^\theta_\text{eff}(x, \tau) = (1-\tau)
\left[U^\theta_\tau(x) - fx\right] + \tau C_\text{prior}$.
It should be emphasized that the correction term $\Delta s^\theta$ arises from the
gradient of the nonlocal integral in the NESS distribution
(Eq.~\eqref{eq:ness_pbc}).
It was found that when the neural network is constrained to output strictly
periodic functions---i.e.,
$U^\theta_\tau(x) = U^\theta_\tau(x + L)$---the integral over one period becomes
independent of $x$, and therefore this term can be neglected.\cite{nagel2025fokker}
This simplification motivates the use of Fourier features as network inputs
(see Sec.~\ref{sec:fourier_features}) and leads to the simplified score
expression used in practice
\begin{equation}
  s^\theta(x_\tau, \tau) \approx
  -\beta (1-\tau)\left[\nabla U^\theta_\tau(x_\tau) -f \right]\label{eq:score_without_correction}\,.
\end{equation}

\subsection{Diffusion on Riemannian Manifolds}

While the formulation above addresses periodic coordinates, estimating free
energies for variables defined on non-Euclidean manifolds, such as angular
coordinates, presents distinct challenges. For a polar angle $\theta$,
geometric factors introduce Jacobian singularities at the poles $\theta = 0,
\pi$, which can lead to numerical instabilities in the effective potential
during training. To address this, diffusion models can be formally extended to
Riemannian manifolds such as SE(3).\cite{corso2022diffdock,yim2023se3}

Here, however, we adopt a more direct approach. By employing the variable
transformation $u = \cos\theta$ and the corresponding force transformation
$f_u = -f_\theta / \sin\theta$, the geometric singularities are absorbed,
yielding a well-defined diffusion process on the interval $u \in [-1, 1]$. The
score function in the original $\theta$ coordinate is then recovered via the
chain rule as
\begin{equation}
  s^\theta(\theta_\tau, \tau) = s^\theta(u_\tau, \tau) \cdot \left|\frac{\mathrm{d}u}{\mathrm{d}\theta}\right|\,.
\end{equation}

\subsection{Extension to Two Dimensions\label{sec:2d_extension}}

The formulation above generalizes naturally to multidimensional collective
variable spaces $\mathbf{x} = (x_1, x_2)$ with periodicities $L_1$ and $L_2$.
In the two-dimensional case, the effective potential becomes
\begin{equation}
  U_\text{eff}(\mathbf{x}, \tau) = (1-\tau) \left[U_\tau(\mathbf{x}) - \mathbf{f}\cdot\mathbf{x}\right] + C_\text{prior}\,,
  \label{eq:U_eff_2d}
\end{equation}
where $\mathbf{f} = (f_1, f_2)$ is the vector of constant driving forces. The
neural network learns a scalar potential $U^\theta_\tau(\mathbf{x})$, and the
score becomes a two-dimensional vector field
\begin{equation}
  s^\theta(\mathbf{x}_\tau, \tau) = -\beta \nabla U^\theta_\text{eff}(\mathbf{x}_\tau, \tau)\,.
  \label{eq:2d_score}
\end{equation}

A key practical consideration is that not all collective variables need to be
externally driven. In the lipid bilayer systems studied here, a constant force
is applied along the membrane normal $z$ (i.e., $f_z \neq 0$), while the
orientational degree of freedom $\theta$ may or may not be biased. When
$f_\theta = 0$, the system still reaches a well-defined NESS because the
$\theta$ coordinate equilibrates conditionally for each value of $z$. The
score-based framework accommodates this naturally by simply setting the
corresponding component of $\mathbf{f}$ to zero. In contrast, for the alanine
dipeptide system, constant forces are applied along both dihedral angles $\phi$
and $\psi$.

Because the neural network learns a scalar energy function, the extension to
two dimensions introduces no structural changes to the architecture beyond
increasing the input dimensionality. The only additional computational cost
arises from the need for sufficient sampling of the two-dimensional NESS
distribution.

\subsection{Training Objective and Regularization\label{sec:training_objective}}

Training the energy-based diffusion model determines the potential energy up to a
diffusion-time-dependent additive constant $C(\tau)$. The full training
objective combines the denoising score matching loss with regularization terms
to ensure physical consistency
\begin{equation}
  \mathcal{L} = \mathcal{L}_\text{DSM} + \lambda_1 \mathcal{L}_\text{reg} + \lambda_2 \mathcal{L}_\text{FP}\,,
  \label{eq:full_loss}
\end{equation}
where $\lambda_1$ and $\lambda_2$ are weighting coefficients. In this work, we
employ the two regularization strategies in a mutually exclusive fashion, setting either $\lambda_1 =
10^{-5}$ (with $\lambda_2=0$) or $\lambda_2 = 10^{-6}$ (with $\lambda_1=0$).


The first regularization term promotes smoothness of the learned potential with
respect to the diffusion time $\tau$\cite{mate2025solvation}
\begin{equation}
  \mathcal{L}_\text{reg} = \mathbb{E}_{x_0\sim p_0, x_\tau\sim p(x_\tau|x_0)} \left[\left\|\partial_\tau U^\theta(x_\tau, \tau)\right\|^2\right]\,.
\end{equation}
While the time-derivative regularization ensures smoothness, it does not
guarantee that the learned potential adheres to the underlying physical dynamics.
Furthermore, in regions of the configuration space that are not sampled,
the network receives no gradient signal from the denoising objective, which can
lead to unphysical extrapolations.

To address these issues, Plainer et al. suggested enforcing consistency with the
Fokker--Planck equation as a physics-informed
regularization, referred to as Fokker--Planck regularization.\cite{plainer2025consistent}
Defining the residual of the diffusion-time-dependent Fokker--Planck equation
\begin{equation}
  R(x, \tau) = -\nabla \cdot J_\tau(x) - \partial_\tau p_\tau(x)\,,
\end{equation}
where the probability current $J_\tau$ and the density $p_\tau$ are
both parameterized through $U^\theta_\tau(x)$ via the NESS ansatz
(Eq.~\eqref{eq:ness_pbc}), the Fokker--Planck regularization reads
\begin{equation}
  \mathcal{L}_\text{FP} = \mathbb{E}_{x_0\sim \mathcal{U}, x_\tau\sim p(x_\tau|x_0)} \left[ \|R(x, \tau)\|^2 \right]\,.
\end{equation}
This physics-informed constraint ensures that the learned potential yields a
self-consistent probability evolution, which is particularly important in
sparsely sampled regions where the denoising objective alone provides
insufficient gradient information (see Sec.~\ref{sec:results}).

Departing from the original formulation, we compute the
expectation value over uniformly sampled initial conditions $x_0\sim\mathcal{U}([0, 1]^d)$ and their
associated noised states $x_\tau$, rather than evaluating the residual solely
on the data distribution $x_0\sim p_0$. This strategy ensures that the physical
constraints are enforced globally across the configuration space, effectively
regularizing the potential even in regions left unexplored by the MD simulation.

To mitigate the computational cost of this additional objective, we employ the
efficient estimator provided by Plainer et al.,
which enables the evaluation of the Fokker--Planck residual using only
first-order spatial derivatives.\cite{plainer2025consistent}

\subsection{Enforcing Symmetries via Fourier Features\label{sec:fourier_features}}

Incorporating system symmetries, particularly periodicity, directly into the
neural network architecture significantly enhances the
performance of Fokker--Planck Score Learning. By constraining the network to
output periodic functions, the correction term $\Delta s^\theta$ in
Eq.~\eqref{eq:score_with_correction} can be neglected, simplifying the training
objective.\cite{nagel2025fokker}
In the one-dimensional case, this is achieved by mapping the spatial coordinate
$x$ with period $L$ to a set of Fourier features
\begin{equation}
  x \to \left\{
    \cos \tfrac{2\pi n}{L}x, \sin \tfrac{2\pi n}{L}x
  \right\}_{n \in \{1, \ldots, N\}}\,, \label{eq:fourier_features_1d}
\end{equation}
with the maximum mode $N$.

This approach can be readily generalized to two dimensions. The most general
form of a smooth, periodic function on a 2D domain can be represented by a
Fourier series expansion in variables $\phi, \psi$ with $2\pi$-periodicity
\begin{equation}
  \begin{aligned}
    f(\phi, \psi) = \smash{\sum_{n, m}}\bigl[&\alpha_{n,m}\cos n \phi\:\cos m \psi\\
      & + \beta_{n,m}\cos n \phi\:\sin m \psi \\
      & + \gamma_{n,m}\sin n \phi\:\cos m \psi \\
    & + \delta_{n,m}\sin n \phi\:\sin m \psi \bigl]\,,\\
  \end{aligned}
\end{equation}
with coefficients defined as
\begin{equation}
  \alpha_{n,m} = \kappa \int_{0}^{2\pi}\mathrm{d}\phi\int_{0}^{2\pi}\mathrm{d}\psi\:f(\phi,\psi)\cos n \phi\:\cos m \psi\,,
\end{equation}
where $\beta_{n,m}$, $\gamma_{n,m}$, and $\delta_{n,m}$ follow analogously with
the respective trigonometric basis functions and the normalization constant
$\kappa$ defined as
\begin{equation}
  4\pi ^2\: \kappa =
  \begin{cases}
    1 & \text{if } n = m = 0\,,\\
    4 & \text{if }n \neq 0 \text{ and } m \neq 0\,,\\
    2 & \text{else}\,.
  \end{cases}
\end{equation}

By utilizing these trigonometric basis functions as input features for the
neural network, boundary conditions are satisfied intrinsically. Specifically,
for physical coordinates $x$ and $y$ with periodicity $L_{x/y}$, the input
transformation becomes
\begin{equation}
  (x, y) \to \left\{
    \begin{array}{l}
      \cos \tfrac{2\pi n}{L_x}x \:\cos \tfrac{2\pi m}{L_y}y, \\
      \sin \tfrac{2\pi n}{L_x}x \:\cos \tfrac{2\pi m}{L_y}y, \\
      \cos \tfrac{2\pi n}{L_x}x \:\sin \tfrac{2\pi m}{L_y}y, \\
      \sin \tfrac{2\pi n}{L_x}x \:\sin \tfrac{2\pi m}{L_y}y
  \end{array}\right\}_{n, m\in \{0, \ldots, N\}}\,.  \label{eq:fourier_features_2d}
\end{equation}
All trivial constant modes (including, e.g., $n=m=0$) are omitted to avoid redundancy.
Beyond periodicity, additional geometric symmetries can be imposed by
selectively including or excluding specific Fourier modes. For instance, mirror
symmetry with respect to $x_1 = 0$, i.e., $U(-x_1, x_2) = U(x_1, x_2)$, is
enforced by retaining only cosine terms in $x_1$.
The specific symmetry constraints employed for each system are detailed in
Sec.~\ref{sec:results}.

\subsection{Molecular Dynamics Simulations\label{sec:md_simulations}}

\subparagraph{Coarse-Grained Lipid Bilayer}
We employed the Martini 3 coarse-grained
force field,\cite{souza21} together with the automated
workflow Martignac for lipid-bilayer simulations.\cite{martignac24}  A symmetric
1-palmitoyl-2-oleoyl-$sn$-glycero-3-phosphocholine (POPC) bilayer was simulated
with a solute modeled by the two Martini beads C1P3. All molecular dynamics
simulations were carried out with GROMACS 2024.3,\cite{berendsen1995gromacs} adhering to
the protocol of Martignac with an integration time step of $\delta t=
\SI{0.02}{\pico\second}$.\cite{martignac24}
Here and throughout this work, time scales for the Martini 3 coarse-grained
model are expressed in picoseconds.
We used a box size of $L_x = L_y = \SI{6}{\nano\meter}$ and $L_z =
\SI{10}{\nano\meter}$, with periodic
boundary conditions in all three dimensions. Each run was equilibrated for
$\SI{200}{\pico\second}$ in the isothermal-isobaric ($NPT$) ensemble at
$T=\SI{298}{\kelvin}$ and $P=\SI{1}{\bar}$, using a stochastic
velocity-rescaling (v-rescale) thermostat\cite{bussi07} and a C-rescale
barostat,\cite{bernetti20} respectively. A semi-isotropic pressure coupling
scheme was employed, with the compressibility in the $z$-direction set to zero
to maintain a constant box height. Subsequent production runs were conducted in
the canonical ($NVT$) ensemble sharing the same temperature and thermostat.
We note that the use of the NVT ensemble was necessitated by GROMACS
requirements to permit biasing across distances exceeding half the simulation
box length.

We performed four independent production runs, each of $\SI{1}{\micro\second}$
duration, applying constant biasing forces $f \in \{4, 6, 8,
10\}\:\si{\kilo\joule\per\mol\per\nano\meter}$. Trajectory frames were recorded
every $\delta t = \SI{0.2}{\pico\second}$ for subsequent analysis.
For each force magnitude, an additional trajectory of $\SI{1}{\micro\second}$
was simulated with a moving harmonic potential applied to the $\theta$ coordinate
(see Sec.~\ref{sec:results} for details).
This additional bias was carried out using the open-source, community-developed
PLUMED library v2.\cite{plumed2019}

For a comparison, we carried out umbrella sampling\cite{torrie1977nonphysical} at a
window spacing of $\delta z = \SI{0.1}{\nano\meter}$ covering the range
$z \in [0, 4.8]\:\si{\nano\meter}$, with \SI{500}{\nano\second} of simulation
per window, for a later reconstruction of the PMF using MBAR.\cite{shirts2008}
Additionally, 50 independent equilibrium MD runs of $\SI{2}{\micro\second}$
each were performed to obtain a converged reference free-energy landscape, where
the first $\SI{500}{\nano\second}$ of each run are discarded for equilibration.

\subparagraph{All-Atom Lipid Bilayer}
For the all-atom system, we simulated ethanol permeation through a POPC lipid
bilayer.
A symmetric bilayer of 100 POPC lipids solvated by 9300 TIP3P water molecules
and containing a single ethanol molecule (parameterized using
CGenFF\cite{vanommeslaeghe2009CGenFF}) is
modeled using the CHARMM36\cite{klauda2010charmm36lipids} force field and
GROMACS 2025.02.\cite{berendsen1995gromacs} Following energy
minimization and a standard six-step CHARMM-GUI\cite{jo2008charmmgui,
feng2023charmmgui,lee2015charmmgui} equilibration protocol
(\SI{1.875}{\nano\second} total) utilizing the v-rescale thermostat and
C-rescale barostat, two NVT production simulations were performed with an
integration time step of \SI{2}{\femto\second} and $T=\SI{308}{\kelvin}$.
Constant-force pulling simulations were performed with $f \in \{20, 25\}\:
\si{\kilo\joule\per\mol\per\nano\meter}$, each with a duration of $t=\SI{60}{\nano\second}$.
The force was applied to the ethanol center of mass along the bilayer normal,
and the molecule's orientation was monitored throughout the simulation.

\subparagraph{Alanine Dipeptide}
We simulated the dynamics of acetyl-alanine-N-methylamide (ACE-ALA-NME)
in explicit solvent. The peptide was
modeled using the AMBER99SB-ILDN\cite{lindorfflarsen2010amberff99sbildn} force field and solvated in a cubic box
containing 622 TIP3P water molecules and one neutralizing $\text{Na}^+/\text{Cl}^-$ ion
pair. All simulations were performed using GROMACS 2025.2.\cite{berendsen1995gromacs}

Following steepest descent energy minimization, the system was equilibrated in
the NPT ensemble for \SI{200}{\pico\second} using the v-rescale thermostat
($T=\SI{298}{\kelvin}$, $\tau_T=\SI{0.1}{\pico\second}$) and C-rescale barostat
($P=\SI{1}{\bar}$, $\tau_P=\SI{2.0}{\pico\second}$, isotropic coupling).

Production simulations were performed in the NVT ensemble at $\SI{298}{\kelvin}$
for \SI{200}{\nano\second} with an integration time step of
\SI{2}{\femto\second}. To enhance sampling of the Ramachandran plane,
constant-force pulling simulations were conducted by applying constant torques
to the backbone torsion angles $\phi$ (C-N-C$_\alpha$-C) and $\psi$
(N-C$_\alpha$-C-N). A set of simulations was performed with force
combinations $f_{\phi, \psi} \in \{(-2, 2), (2, -2), (0, 2), (0, -2)\} \: \si{\kilo\joule\per\mol\per\radian}$.
As a reference, 12 independent equilibrium MD runs of
\SI{2.4}{\micro\second} each were performed, totaling \SI{28.8}{\micro\second}.

\subsection{Neural Network Architecture and Training\label{sec:nn_training}}

The diffusion model was implemented using the JAX machine learning
framework,\cite{jax} the Flax neural network
library,\cite{flax} and the Optax optimization framework.\cite{optax}

\subparagraph{Noise Schedule}
An exponential noise schedule was employed,\cite{song2020score}
\begin{equation}
  \sigma_\tau = \sigma_\text{min}^{1-\tau} \sigma_\text{max}^\tau \,,
\end{equation}
where $\sigma_{\text{min}}$ and $\sigma_{\text{max}}$ denote the minimum and
maximum noise levels, respectively. For all systems,
$\sigma_\text{max}$ was set to $L/2$, ensuring
that the noising process reaches the uniform latent distribution. The minimum
noise level was set to $\sigma_\text{min} = 0.005$ for the lipid systems to
ensure high spatial resolution in the learned potential energy function, while
for alanine dipeptide we used a larger minimal noise $\sigma_\text{min}=0.01$
to restrict overfitting.

\subparagraph{Optimization}
To ensure consistency, all models were trained for $50$ epochs using the AdamW
optimizer\cite{loshchilov17} with a mini-batch size of $512$. The learning rate
followed an initial linear warm-up from $5\times 10^{-7}$ to $5 \times 10^{-3}$
over the first 5 epochs, followed by a cosine decay to $5\times 10^{-7}$ for the
remaining epochs.
All available trajectory frames were used for training,
corresponding to a sampling interval of $\Delta t = \SI{0.02}{\pico\second}$
for alanine dipeptide, and $\Delta t = \SI{0.2}{\pico\second}$ and $\Delta
t=\SI{0.06}{\pico\second}$ for the coarse-grained and all-atom lipid bilayer
systems, respectively.

\subparagraph{Architecture}
The neural network consisted of a fully connected architecture with the Swish
activation function\cite{ramachandran17} to ensure differentiability. For the
lipid systems (coarse-grained and all-atom), the network comprised five hidden
layers of 64 neurons each, while for alanine dipeptide, 3 layers of 48
neurons each were used.

\section{Results and Discussion}\label{sec:results}

\subsection{Alanine Dipeptide}

\begin{figure}[tb]
  \centering
  \includegraphics{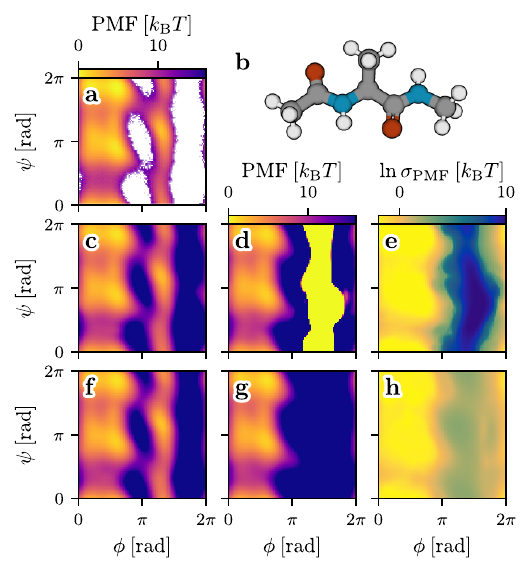}
  \caption{
    Free-energy estimation of alanine dipeptide in water:
    (a) Reference Ramachandran plot of the free-energy landscape as a function
    of the dihedral angles $\phi$ and $\psi$.
    (b) Depiction of alanine dipeptide.
    (c--h) Reconstructed free-energy landscapes from \SI{23}{\nano\second} of
    non-equilibrium MD data, with (c, f) and without (d, g) sampling of the $\alpha_L$ region.
    The upper row corresponds to the simple energy regularization and the lower row to the Fokker--Planck regularization.
    (e+h) The standard deviation across 50 independent runs for the simple
    energy regularization and the Fokker--Planck regularization, respectively.
  }
  \label{fig:diala_2D_results}
\end{figure}
We begin with alanine dipeptide solvated in water, a system characterized by
two intrinsically periodic collective variables: the backbone dihedral angles
$\phi$ and $\psi$. This application represents a direct extension of the
one-dimensional FPSL framework\cite{nagel2025fokker} to two dimensions, as
both coordinates are periodic and can be driven by constant biasing torques.
In contrast to the lipid bilayer system, where the periodicity of the system
results from the periodic boundary conditions of the simulation box, here the
periodicity arises from the molecular structure itself---specifically the
dihedral angles.

As a reference, we performed 12 independent equilibrium MD simulations of
\SI{2.4}{\micro\second} each, totaling \SI{28.8}{\micro\second}. The resulting
free-energy landscape as a function of the dihedral angles $\phi$ and $\psi$ is
shown in Fig.~\subref{fig:diala_2D_results}{a}. For the non-equilibrium MD
simulations, we applied constant biasing forces
$f_{\phi,\psi} = (0,\pm 2)\,\si{\kilo\joule\per\mol\per\radian}$ and
$f_{\phi,\psi} = (\mp2,\pm 2)\,\si{\kilo\joule\per\mol\per\radian}$, simulating
\SI{200}{\nano\second} for each force combination.

Although both dihedral angles are periodic, the constant biasing torques are less
effective at enhancing sampling than in the membrane systems discussed below.
The steric clash near $\phi \approx 2\pi$
effectively confines $\phi$, making it behave akin to a
non-periodic coordinate. While this limitation could be mitigated by introducing
a time-dependent biasing potential, we purposefully retain this setup to
demonstrate the impact of insufficient sampling of relevant degrees of freedom on
the reconstruction of the free-energy landscape.

Figure~\subref{fig:diala_2D_results}{c+d} depicts the reconstructed free-energy
landscapes derived from \SI{11.5}{\nano\second} of MD simulation data for each
force combination, $f_{\phi,\psi} = (\mp2,\pm 2)\,\si{\kilo\joule\per\mol\per\radian}$,
amounting to a cumulative sampling time of \SI{23}{\nano\second}.
In Fig.~\subref{fig:diala_2D_results}{c}
we select an individual run where the left-handed $\alpha$-helix region ($\alpha_L$) was sampled,
whereas in Fig.~\subref{fig:diala_2D_results}{d} we show a run where
this region remained unexplored. While both
reconstructions successfully recover the main free-energy basins ($\phi < \pi$),
the lack of sampling in the $\alpha_L$ region in Fig.~\subref{fig:diala_2D_results}{d} results in an arbitrary
network output. This behavior arises because the neural network is never exposed
to samples in this region during training, and consequently, the objective function
provides no gradient to guide the potential energy surface prediction in this area.

To address this issue, we employ the Fokker--Planck
regularization,\cite{plainer2025consistent} as outlined in section
\ref{sec:theory}. This approach imposes a physics-informed constraint on the learned
energy landscape, ensuring consistency with the stationarity condition of the
Fokker--Planck equation. Incorporating this regularization leads to a significant
improvement in the reconstruction of the unsampled $\alpha_L$ region, as shown in
Fig.~\subref{fig:diala_2D_results}{f+g}. With this constraint, FPSL robustly
identifies the unsampled area as a high-energy state. Notably, when $\alpha_L$ is
adequately sampled, both regularization schemes yield virtually identical
free-energy landscapes.

To quantify these observations, Fig.~\subref{fig:diala_2D_results}{e+h} displays
the logarithm of the standard deviation across 50 independent training runs for
both regularization methods. It is evident that while both approaches converge clearly
within the sampled regions, the Fokker--Planck regularization exhibits a
significantly reduced standard deviation in the unsampled $\alpha_L$ region. In
contrast, the simple energy regularization is prone to numerical artifacts and
high variance in the absence of data.

\begin{figure}[tb]
  \centering
  \includegraphics{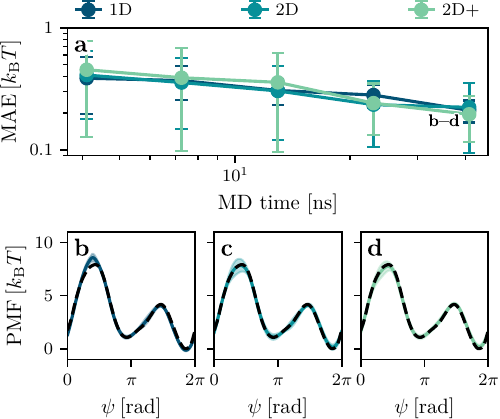}
  \caption{
    Performance evaluation of Fokker--Planck Score Learning (FPSL) for reconstructing
    the one-dimensional free-energy profile of alanine dipeptide along the $\psi$ dihedral angle.
    (a) Mean absolute error (MAE) as a function of the aggregate MD simulation time. The comparison includes:
    FPSL trained solely on $\psi$ (1D);
    FPSL trained on the joint space $(\phi, \psi)$ with biasing applied only along $\phi$ (2D); and
    FPSL trained on $(\phi, \psi)$ with biasing applied along both $\phi$ and $\psi$ (2D+).
    For the 2D approaches, the Fokker--Planck regularization scheme is employed, and the corresponding
    1D profile is derived by marginalizing over $\phi$.
    (b--d) Comparison of the free-energy profiles reconstructed by FPSL using
    approximately \SI{41}{\nano\second} of non-equilibrium MD data for the 1D (b),
    2D (c), and 2D+ (d) methodologies.
  }
  \label{fig:diala_mae_results}
\end{figure}

Finally, we evaluate the performance of FPSL in reconstructing the one-dimensional
free-energy profile along the $\psi$ dihedral angle. Because
the orthogonal coordinate $\phi$ exhibits slow relaxation dynamics compared to $\psi$,
the Fokker--Planck regularization scheme is employed
throughout the following analysis to ensure robust learning of the energy landscape.

Figure~\subref{fig:diala_mae_results}{a} presents the convergence of the MAE as a function
of the accumulated MD simulation time. Three distinct approaches are compared:
($i$) direct learning of the 1D free-energy profile along $\psi$ (1D);
($ii$) learning the 2D free-energy landscape in $(\phi, \psi)$ followed by marginalization over $\phi$ (2D);
and ($iii$) learning the 2D landscape while additionally biasing the $\phi$ coordinate (2D+).
Although all three methods converge to a comparable error of approximately $0.2\:k_\text{B}T$
after an aggregate simulation time of \SI{41}{\nano\second}, the 2D/2D+
approaches demonstrate slightly faster convergence behavior.
However, it is observed that the standard deviation across 50 independent training runs
is notably higher for the 2D and 2D+ approaches compared to the 1D case. This increased
variance can likely be attributed to the sparse sampling of the slow $\phi$ degree of freedom.

The reconstructed free-energy profiles obtained from approximately \SI{41}{\nano\second}
of MD simulation data are visualized in Fig.~\subref{fig:diala_mae_results}{b--d} for the 1D,
2D, and 2D+ approaches, respectively. While all three methods successfully capture
the principal features of the reference profile, the 1D approach exhibits a systematic
bias in the barrier region near $\psi \approx \pi / 2$. This deviation is identified
as a consequence of insufficient sampling in the $\alpha_L$ region (cf. Fig.~\subref{fig:diala_2D_results}{a}).
Learning the full 2D landscape and subsequently marginalizing over $\phi$ (2D) effectively
mitigates this bias.
It is worth noting that the left-handed $\alpha$-helix ($\alpha_L$) constitutes only a
small fraction of the total equilibrium probability mass. Consequently, neglecting this region
in our reference equilibrium simulations introduces only a minor deviation in the
1D free-energy profile along $\psi$, resulting in a baseline error of approximately
$0.18\:k_\text{B}T$, which is comparable to the accuracy achieved by our FPSL results.

\subsection{Coarse-Grained Lipid Bilayer}
\begin{figure}[tb]
  \centering
  \includegraphics{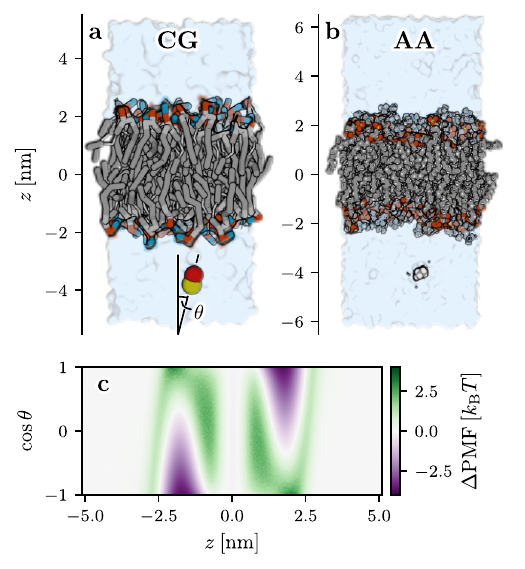}
  \caption{Molecular systems studied:
    (a) Coarse-grained Martini 3 model of a C1P3 molecule permeating a POPC lipid bilayer.
    (b) All-atom model of an ethanol molecule permeating a POPC lipid bilayer.
    (c) Free-energy landscape of C1P3 permeating a POPC lipid bilayer, shown in
    panel (a), as a function of the normal distance $z$ from the bilayer
    midplane and $\cos\theta$, where $\theta$ is the polar angle of the molecule.
    \label{fig:ref}
  }
\end{figure}

We next turn to the lipid bilayer, a fundamental structural
component of biological membranes whose free-energy landscape governs critical
phenomena such as solute permeation.
In contrast to the alanine dipeptide system, where both collective variables are
periodic dihedral angles, the lipid bilayer introduces a qualitatively different
challenge: one of the two relevant collective variables---the polar angle
$\theta$ characterizing the solute orientation---is intrinsically non-periodic.
This extends the FPSL framework beyond purely periodic CV spaces and
demonstrates its flexibility in handling mixed coordinate types.
In previous work, we introduced Fokker--Planck Score Learning (FPSL) as a method to
reconstruct free-energy landscapes from non-equilibrium steady-state
simulations.\cite{nagel2025fokker} Therein, the efficiency of the method relative
to conventional umbrella sampling combined with the multistate Bennett acceptance
ratio (MBAR) estimator was established by reconstructing the
one-dimensional free-energy profile of a single Martini bead permeating a
1-palmitoyl-2-oleoyl-sn-glycero-3-phosphocholine (POPC) bilayer.

In the present study, we extend FPSL to systems of increased complexity to demonstrate:
($i$) its robustness in reconstructing one-dimensional free-energy profiles even
when relevant orthogonal degrees of freedom are marginalized, and ($ii$) its capacity
to reconstruct full two-dimensional free-energy landscapes with negligible
computational overhead compared to the one-dimensional case.
Specifically, we employ the coarse-grained Martini force field to investigate
the permeation of a solute composed of C1 and P3 beads through a POPC bilayer,
as illustrated in Fig.~\subref{fig:ref}{a}.
A detailed description of the molecular dynamics simulation parameters is
provided in Sec.~\ref{sec:md_simulations}.

The system dynamics are primarily characterized by two collective variables:
($i$) the normal distance of the center of mass of C1P3 from the bilayer midplane,
denoted by $z$, and ($ii$) the polar angle $\theta$ between the
bilayer normal and the vector connecting the C1 and P3 beads (see Fig.~\subref{fig:ref}{a}).
The regions $z<0$ and $z>0$ correspond to the lower and upper leaflets,
respectively. Given the amphiphilic nature of the solute (hydrophobic C1, hydrophilic P3),
the polar angle $\theta$ plays a significant role in determining the insertion
free energy. The resulting two-dimensional free-energy landscape as a function of
$z$ and $\cos\theta$ is depicted in Fig.~\subref{fig:ref}{c}. It features two
symmetric minima at $(z, \cos\theta) \approx (\pm\SI{1.8}{\nano\meter}, \pm 1)$,
corresponding to the preferred orientation of the molecule in each leaflet, with
the hydrophobic C1 bead oriented toward the bilayer core.

We begin by reconstructing the 1D free-energy profile along the $z$-coordinate,
a metric commonly used to characterize solute permeation through lipid
bilayers. Here, we compare the performance of FPSL when learning the 1D FEL directly
versus learning the 2D FEL along $z$ and $\theta$ and subsequently marginalizing
over $\theta$. To sample the non-equilibrium steady state, we employ biased MD
simulations by applying constant forces to the C1P3 molecule along the $z$-direction
of $f\in\{4, 6, 8, 10\}\:\si{\kilo\joule\per\mole\per\nano\meter}$, each for a
duration of $t=\SI{1}{\micro\second}$.
For each force magnitude, an additional trajectory of $\SI{1}{\micro\second}$
is simulated with a moving harmonic potential applied to the $\theta$ coordinate,
a technique commonly utilized in steered MD biasing
\begin{equation}
  f_\theta(t) = \kappa \left[ \frac{\pi}{2} \left( 1 + \sin \frac{2\pi t}{\tau} \right) - \theta \right]
\end{equation}
with $\kappa = \SI{0.5}{\kilo\joule\per\mole\per\radian\squared}$ and $\tau =
\SI{39.813}{\pico\second}$. It is important to note that, as $\theta$ is not
periodic, a constant force would bias the system towards $\theta = \pi$ rather than
enhancing sampling. Furthermore, the period is chosen such that it is not a multiple
of the integration step, ensuring adequate coverage.

\begin{figure}[tb]
  \centering
  \includegraphics{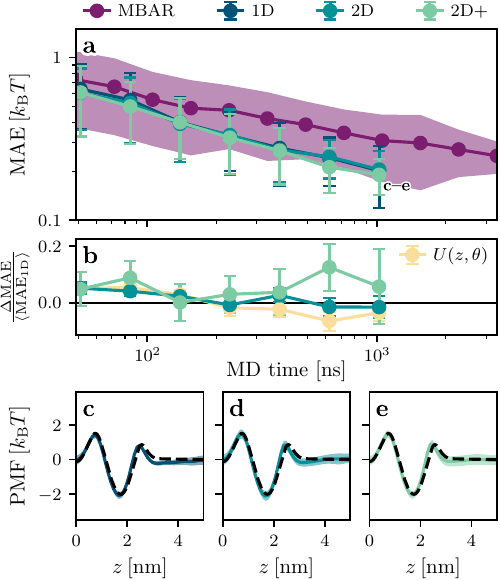}
  \caption{
    Performance evaluation of Fokker--Planck Score Learning (FPSL) for reconstructing
    the free-energy profile of C1P3 permeation through a POPC lipid bilayer.
    (a) Mean absolute error (MAE) as a function of aggregate MD simulation time. Comparison includes:
    FPSL trained on $z$ (1D);
    FPSL trained on $(z, \cos\theta)$ with biasing along $z$ (2D);
    FPSL trained on $(z, \cos\theta)$ with biasing along both $z$ and $\theta$ (2D+);
    and umbrella sampling with MBAR (purple).
    For the 2D methods, the 1D profile is obtained by marginalizing over $\cos\theta$.
    (b) Convergence improvement ($\Delta \text{MAE}$) achieved by learning the
    full 2D landscape (and subsequently marginalizing) compared to learning the
    1D landscape directly. The yellow curve represents the 2D reconstruction
    using $\theta$ instead of $\cos\theta$.
    (c--e) Comparison of free-energy profiles reconstructed by FPSL using
    approximately $\SI{1}{\micro\second}$ of MD simulation data.
  }
  \label{fig:c1p3_1d_mae}
\end{figure}

As a reference, we utilize equilibrium simulations with a total duration of
$\SI{97.5}{\micro\second}$, obtained from 50 independent MD runs. From these,
the free-energy profile along $z$ is derived via Fourier expansion of the binned data.
This profile, shown in Fig.~\subref{fig:c1p3_1d_mae}{c--d} as a black dashed line,
serves as a benchmark for evaluating the performance of our FPSL method. We quantify
accuracy via the mean absolute error (MAE) of the reconstructed free-energy profiles
$U(z)$ relative to this reference
\begin{equation}
  \text{MAE} = \left\langle \left| U(z) - U_\text{ref}(z) \right| \right\rangle_z\,.
\end{equation}

Following the methodology established in our previous work,\cite{nagel2025fokker}
we utilize Fourier features to parametrize the neural network representing the potential
energy function $U^\theta(x)$. For the 1D free-energy profile along $z$, we exploit
the system's symmetry with respect to the bilayer midplane, i.e., $U(-z) = U(z)$.
This symmetry is enforced by restricting the network parametrization to cosine
Fourier modes, utilizing the first $N=8$ modes.
For the 2D free-energy landscape along $z$ and $\theta$, we leverage two geometric
properties: ($i$) the central symmetry of the landscape around $(0, \pi/2)$, implying
$U(-z, \theta) = U(z, \pi - \theta)$, and ($ii$) the reflective boundary conditions
of the $\theta$ coordinate at $\theta = 0$ and $\theta = \pi$. The latter is enforced
by mapping $\theta$ to a $2\pi$-periodic coordinate described solely by even (cosine)
modes. We parametrize the neural network by retaining all Fourier modes (see section
\ref{sec:fourier_features}) that satisfy the requested symmetries. Introducing
the shifted coordinate $\vartheta = \theta - \pi / 2$, this yields
\begin{align}
  \left\{
    \cos \tfrac{2\pi n}{L} z\: \cos \:2m \vartheta,
    \sin \tfrac{2\pi n}{L} z\: \sin \:(2m+1) \vartheta
  \right\}_{n,m} \label{eq:fourier_feature_parametrization}
\end{align}
where the number of modes is limited to $N=4$, and the factor of 2 in the $\vartheta$
modes ensures evenness in $\theta$. The trivial mode $n=m=0$ is excluded.

To facilitate the training of the 2D free-energy landscape, the potential is learned
as a function of $U(z, \cos\theta)$ rather than $U(z, \theta)$. This transformation
results in a uniform FEL in the water phase, thereby improving training efficiency.
Correspondingly, the forces in $\theta$ are scaled by $f_{\cos\theta} = -f_\theta / \sin\theta$
to account for the change of variables.

To evaluate method performance, the neural network was trained using datasets of
varying trajectory lengths, $t \in [12.8, 256]\:\si{\nano\second}$ for each force.
These datasets comprised segments randomly selected from each constant-force simulation.
To assess the uncertainty of the reconstructed profiles, this procedure was repeated
for 50 independent random selections and subsequent network trainings for each trajectory
length.
For comparison with conventional umbrella sampling--based MBAR, simulations were
conducted with a window spacing of $\delta z = \SI{0.1}{\nano\meter}$ over the range
$z \in [0, 4.8]\:\si{\nano\meter}$. Each window was simulated for $t=\SI{500}{\nano\second}$,
resulting in a total simulation time of \SI{24.5}{\micro\second}. To estimate uncertainty,
MBAR was applied to trajectory segments of lengths $t \in [1, 68]\:\si{\nano\second}$
from each window.

Figure~\subref{fig:c1p3_1d_mae}{a} displays the MAE of MBAR versus FPSL using only
$z$ as input (1D), $z$ and $\theta$ as input (2D), and additionally biasing $\theta$
(2D+) as a function of the accumulated MD simulation time.
All three FPSL variants are observed to converge significantly faster than MBAR.
Consequently, we compare in Fig.~\subref{fig:c1p3_1d_mae}{b} the relative change
in MAE between the 1D and 2D approaches, defined as
\begin{equation}
  \Delta \text{MAE} = \langle \text{MAE}_\text{1D} - \text{MAE}_\text{2D} \rangle\,.
\end{equation}
It is found that learning the full 2D landscape and subsequently marginalizing
over $\cos\theta$ (2D) yields a comparable MAE to learning the 1D landscape directly (1D).
Thus, reconstructing the full 2D FEL does not introduce additional uncertainty into
the marginalized 1D free-energy profile.
However, when $\theta$ is additionally biased (2D+), a slight improvement in the
MAE is observed. The magnitude of this improvement is small, presumably because
the sampling along $\theta$ is already sufficient when biasing only along $z$, due to the
significantly faster relaxation time of $\theta$ compared to $z$. For the same reason,
neglecting $\theta$ in the 1D case (or in MBAR) does not substantially impact performance.

It should be noted that when dealing with coordinates lacking a uniform measure,
such as polar angles, it can be advantageous to learn the FEL using $\cos\theta$
rather than $\theta$ directly, as this results in a uniform measure in the latent space
(see Fig.~\subref{fig:ref}{c}). Indeed, we find that employing $\cos\theta$ leads to
a minor improvement, as shown in Fig.~\subref{fig:c1p3_1d_mae}{b}.

To visualize these findings, Fig.~\subref{fig:c1p3_1d_mae}{c--e} present the mean
and variance of the reconstructed free-energy profiles using approximately
$\SI{1}{\micro\second}$ of MD simulation data for the 1D, 2D, and 2D+ approaches,
respectively.
All three FPSL approaches yield profiles in good agreement with the reference (black dashed line),
with the 2D+ approach exhibiting the closest correspondence.
While the 1D approach displays the smallest variance, it exhibits a slight bias in the
barrier region at $z\approx\SI{2.5}{\nano\meter}$, which is mitigated when learning
the full 2D landscape. This bias likely arises because, under strong pulling forces,
the molecule lacks sufficient time to reorient from $\theta \approx 0$ to $\theta \approx \pi$
as it exits the membrane. Consequently, the system attempts an insertion in an
unfavorable orientation, introducing a systematic bias in the calculated free-energy profile.

\begin{figure}[tb]
  \centering
  \includegraphics{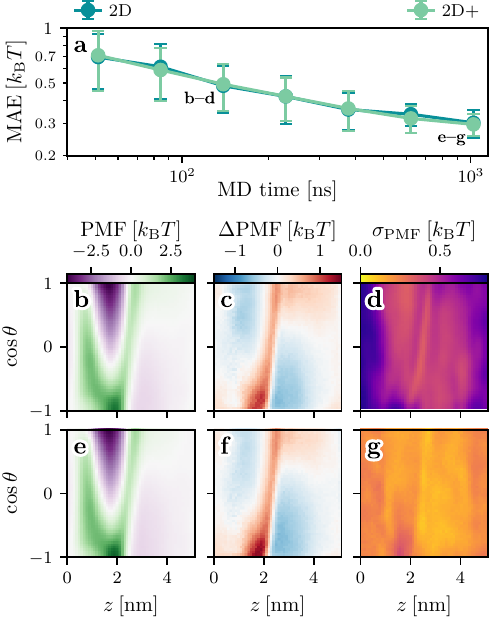}
  \caption{
    Performance of Fokker--Planck Score Learning (FPSL) in reconstructing the 2D
    free-energy landscape of C1P3 permeating a POPC lipid bilayer. (a) Convergence of the
    2D mean absolute error (MAE) as a function of aggregate MD simulation time, comparing
    FPSL trained on $(z, \cos\theta)$ with biasing applied only along $z$ (2D) versus
    biasing applied along both $z$ and $\theta$ (2D+). (b--g) Analysis of the
    landscapes learned by the 2D method (bias on $z$ only) using approximately
    $\SI{139}{\nano\second}$ (top row) and $\SI{1}{\micro\second}$ (bottom row) of data.
    Panels show the (b, e) average reconstructed FEL, (c, f)
    difference to the reference FEL, and (d, g) the standard deviation across 50
    independent training runs.
  }
  \label{fig:c1p3_2d_mae}
\end{figure}

Subsequently, we assess the performance of FPSL in reconstructing the full 2D
free-energy landscape as a function of $z$ and $\cos\theta$.
We compare the performance of FPSL when biasing solely along $z$ (2D) versus
biasing along both $z$ and $\theta$ (2D+), as shown in Fig.~\subref{fig:c1p3_2d_mae}{a}.
Notably, these results utilize the same training data as the 1D analysis above, but
without marginalization over $\cos\theta$.
Both methods are found to converge to virtually identical MAE values, reaching
approximately $0.32\:k_\text{B}T$ after $\SI{1}{\micro\second}$ of
aggregate MD simulation time, compared to approximately $0.24\:k_\text{B}T$ in
the 1D case.

In Fig.~\subref{fig:c1p3_2d_mae}{b--g}, we analyze the reconstructed 2D free-energy
landscape utilizing the 2D method (biasing only along $z$) after approximately
$\SI{139}{\nano\second}$ (top row) and $\SI{1}{\micro\second}$ (bottom row) of
MD simulation data, averaged over 50 independent training runs. In both instances,
the principal features of the free-energy landscape are accurately reconstructed,
including the minimum at $(z, \cos\theta) \approx (\SI{1.8}{\nano\meter}, 1)$,
the U-shaped barrier region surrounding it, and the uniform free energy in the water
phase (see \subref{fig:c1p3_2d_mae}{b+e}).
However, upon comparison with the reference FEL, FPSL exhibits a small systematic
bias, specifically an overestimation of the barrier height (indicated in dark red),
see \subref{fig:c1p3_2d_mae}{c+f}. This deviation is presumably attributable to the
sparse sampling of this high-energy region in the non-equilibrium steady-state data,
which results in higher uncertainty in the reconstructed landscape.
Examination of the standard deviation across the 50 independent training runs
(see \subref{fig:c1p3_2d_mae}{d+g}) reveals uniform convergence across the
landscape as sampling increases.

\subsection{All-Atom Lipid Bilayer}

\begin{figure}[tb]
  \centering
  \includegraphics{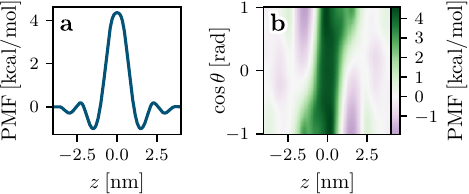}
  \caption{
    Free-energy landscape of ethanol permeating a POPC lipid bilayer modeled at
    the all-atom level (see Fig.~\subref{fig:ref}{b}), reconstructed using
    Fokker--Planck Score Learning (FPSL). The landscape is depicted as a
    function of the normal distance $z$ from the bilayer midplane and the cosine
    of the polar angle, $\cos\theta$.
    (a) Marginalized 1D free-energy profile obtained by integrating over
    $\cos\theta$, and (b) the full reconstructed 2D landscape.
  }
  \label{fig:etoh_results}
\end{figure}

Having established and benchmarked the performance of FPSL on a coarse-grained model
with a non-periodic orientational degree of freedom,
we now demonstrate its applicability at all-atom resolution: the permeation of ethanol
through a POPC lipid bilayer (see Fig.~\ref{fig:ref}).
This system has been previously investigated by Comer et al., who reported a
2D free-energy landscape.\cite{comer2014diffusive}
Consistent with their work, we characterize the system using two collective
variables: ($i$) the normal distance $z$ of the ethanol center of mass from the
bilayer midplane, and ($ii$) the polar angle $\theta$ between the $z$-axis and the
vector $r_\text{Et} - r_\text{OH}$, where $r_\text{OH}$ and $r_\text{Et}$ denote the
centers of mass of the hydroxyl and ethyl groups, respectively.

We perform constant-force pulling simulations with $f \in \{20, 25\}\:
\si{\kilo\joule\per\mol\per\nano\meter}$, each with a duration of $t=\SI{60}{\nano\second}$,
resulting in 10 and 14 membrane permeation events, respectively.
Analogous to the coarse-grained case, we enforce point symmetry around
$(z, \theta) = (0, \pi/2)$ as well as reflective boundary conditions at $\theta = 0$
and $\theta = \pi$ by employing the Fourier features defined in
Eq.~\eqref{eq:fourier_feature_parametrization} with $N=4$.
Using FPSL, we reconstruct the 2D free-energy landscape as a function of $z$ and
$\cos\theta$, shown in Fig.~\subref{fig:etoh_results}{b}.

As described by Comer et al., the free-energy landscape exhibits distinct features:
($i$) two minima at $(z, \cos\theta) \approx (\pm\SI{1.5}{\nano\meter}, \pm 1)$,
corresponding to an orientation where the hydroxyl group points toward the water phase
while the ethyl group is buried in the hydrophobic core; and ($ii$) an S-shaped barrier region
spanning approximately $\SI{2}{\nano\meter}$.
Regarding the barrier height, we obtain a value of approximately \SI{4}{\kilo\cal\per\mol},
which is in reasonable agreement with the value of $\approx \SI{3.0}{\kilo\cal\per\mol}$
reported by Comer et al.
For the depth of the minima, we observe slightly larger values (\SI{1.8}{\kilo\cal\per\mol})
compared to the reference (\SI{1.0}{\kilo\cal\per\mol}). This discrepancy can be attributed
to the different coordinate representations: Comer et al. computed $U(z, \theta)$, which
implicitly includes the Jacobian term $-k_\text{B}T\ln \sin \theta$, resulting in
shallower minima near the poles ($\theta = 0, \pi$).
It should be noted that while our result, $U(z, \cos\theta)$,
yields a uniform free energy in the bulk water phase, which corresponds to the
expected physical behavior, the uniformity observed in the reference work
in the bulk water phase suggests potential sampling or convergence
limitations in the original study.

The 1D free-energy profile for ethanol permeation has also been reported by Ghorbani
et al.\cite{ghorbani2020molecular}. They observed that the permeation barrier increases
with decreasing ethanol concentration, rising from $3.3 \pm 1.0\:k_\text{B}T$ to
$6.3\pm0.1\:k_\text{B}T$ as the concentration drops from \SI{18}{\mol\percent} to
\SI{1}{\mol\percent}. Our simulation of a single ethanol molecule corresponds to
the dilute limit ($< \SI{0.01}{\mol \percent}$). We find a barrier of approximately
\SI{4.2}{\kilo\cal\per\mol} (approx. $6.9\:k_\text{B}T$), which is in excellent agreement
with their results at low concentration.

Comer et al. employed adaptive biasing force (ABF) simulations to reconstruct
the 2D free-energy landscape. Although the computational cost for the 2D surface
was not explicitly reported, a subsequent study noted that generating 1D free-energy
profiles for ethanol permeation across various bilayers required approximately
$2\text{--}4\:\si{\micro\second}$ of sampling.\cite{tse2019affordable}
In contrast, Ghorbani et al. reported that approximately \SI{0.5}{\micro\second}
was sufficient to reconstruct the 1D free-energy profile of ethanol in POPC
using maximum likelihood estimation.\cite{Kraemer2020membrane, ghorbani2020molecular}
Our FPSL approach enables the reconstruction of the full 2D free-energy
landscape with as little as \SI{120}{\nano\second} of MD simulation,
representing a significant reduction in computational cost compared to these
established methods.
Specifically, this corresponds to more than an order-of-magnitude speedup
relative to ABF-based approaches and approximately a fourfold reduction
compared to maximum likelihood estimation, while simultaneously providing the
full two-dimensional landscape rather than only a one-dimensional projection.

\section{Conclusion\label{sec:conclusion}}

In this work, we have extended the Fokker--Planck Score Learning (FPSL)
framework to reconstruct multidimensional free-energy landscapes from
non-equilibrium molecular dynamics simulations. By incorporating orthogonal
degrees of freedom into the learning process, we demonstrated that FPSL can
recover full two-dimensional landscapes with negligible computational overhead
compared to one-dimensional profiles. We validated this approach on systems of
varying complexity, from the conformational dynamics of alanine dipeptide---where
both collective variables are periodic dihedral angles---to coarse-grained and
all-atom lipid bilayer permeation, where the orientational degree of freedom
$\theta$ is intrinsically non-periodic.
Remarkably, for the all-atom lipid bilayer, FPSL reconstructs the full 2D
free-energy landscape from only \SI{120}{\nano\second} of MD simulation---around
an order of magnitude less than ABF-based approaches and maximum likelihood
estimation, both of which yield only one-dimensional profiles.

Importantly, because FPSL learns a smooth score function rather than relying on
histogram-based densities like most established methods, incorporating a second
collective variable incurs minimal additional cost during the learning phase.
Our results show that while direct one-dimensional learning is often already
accurate, explicitly learning the joint free-energy landscape can prevent
small systematic biases that can arise when orthogonal degrees of freedom relax
slowly or are strongly perturbed by external forces. Specifically, in the case
of alanine dipeptide, resolving the $\phi$ degree of freedom corrected
deviations in the barrier region. Furthermore, we showed that appropriate
coordinate transformations, such as learning in $\cos\theta$ space, can
significantly enhance training efficiency by ensuring a more uniform probability
measure. Notably, the framework handles also non-periodic collective variables
naturally, requiring only that their boundary conditions
allow for mapping to a periodic domain.
Furthermore, we demonstrated the efficacy of the Fokker--Planck regularization
in systems with complex energy landscapes, such as alanine dipeptide. While
standard score matching can yield arbitrary potentials in unsampled regions,
enforcing consistency with the stationary solution of the Fokker--Planck
equation ensures that high-energy states are robustly identified even in the
absence of direct sampling. This physics-informed constraint effectively guides
the neural network in data-sparse regimes, dramatically reducing the variance of
the reconstructed landscape and ensuring physically meaningful predictions.
Making FPSL numerically robust to sparsely sampled regions is a crucial step towards
estimating higher-dimensional free-energy landscapes, $D > 2$.

The ability to efficiently reconstruct multidimensional free-energy surfaces
opens new avenues for studying complex biomolecular processes. Future work will
focus on scaling FPSL to even higher-dimensional reaction coordinates and
integrating it with advanced sampling strategies to explore rare events. We
anticipate that the combination of data-efficient score-based learning with
robust physics-informed regularization will become a powerful tool for
characterizing the thermodynamics and kinetics of molecular systems.

%% file: main.bbl
\begin{thebibliography}{56}%
\makeatletter
\providecommand \@ifxundefined [1]{%
 \@ifx{#1\undefined}
}%
\providecommand \@ifnum [1]{%
 \ifnum #1\expandafter \@firstoftwo
 \else \expandafter \@secondoftwo
 \fi
}%
\providecommand \@ifx [1]{%
 \ifx #1\expandafter \@firstoftwo
 \else \expandafter \@secondoftwo
 \fi
}%
\providecommand \natexlab [1]{#1}%
\providecommand \enquote  [1]{``#1''}%
\providecommand \bibnamefont  [1]{#1}%
\providecommand \bibfnamefont [1]{#1}%
\providecommand \citenamefont [1]{#1}%
\providecommand \href@noop [0]{\@secondoftwo}%
\providecommand \href [0]{\begingroup \@sanitize@url \@href}%
\providecommand \@href[1]{\@@startlink{#1}\@@href}%
\providecommand \@@href[1]{\endgroup#1\@@endlink}%
\providecommand \@sanitize@url [0]{\catcode `\\12\catcode `\$12\catcode `\&12\catcode `\#12\catcode `\^12\catcode `\_12\catcode `\%12\relax}%
\providecommand \@@startlink[1]{}%
\providecommand \@@endlink[0]{}%
\providecommand \url  [0]{\begingroup\@sanitize@url \@url }%
\providecommand \@url [1]{\endgroup\@href {#1}{\urlprefix }}%
\providecommand \urlprefix  [0]{URL }%
\providecommand \Eprint [0]{\href }%
\providecommand \doibase [0]{https://doi.org/}%
\providecommand \selectlanguage [0]{\@gobble}%
\providecommand \bibinfo  [0]{\@secondoftwo}%
\providecommand \bibfield  [0]{\@secondoftwo}%
\providecommand \translation [1]{[#1]}%
\providecommand \BibitemOpen [0]{}%
\providecommand \bibitemStop [0]{}%
\providecommand \bibitemNoStop [0]{.\EOS\space}%
\providecommand \EOS [0]{\spacefactor3000\relax}%
\providecommand \BibitemShut  [1]{\csname bibitem#1\endcsname}%
\let\auto@bib@innerbib\@empty
\bibitem [{\citenamefont {Chipot}\ and\ \citenamefont {Pohorille}(2007)}]{chipot2007free}%
  \BibitemOpen
  \bibfield  {author} {\bibinfo {author} {\bibfnamefont {C.}~\bibnamefont {Chipot}}\ and\ \bibinfo {author} {\bibfnamefont {A.}~\bibnamefont {Pohorille}},\ }\href@noop {} {\emph {\bibinfo {title} {Free energy calculations}}},\ Vol.~\bibinfo {volume} {86}\ (\bibinfo  {publisher} {Springer},\ \bibinfo {year} {2007})\BibitemShut {NoStop}%
\bibitem [{\citenamefont {Hansen}\ and\ \citenamefont {Van~Gunsteren}(2014)}]{hansen2014practical}%
  \BibitemOpen
  \bibfield  {author} {\bibinfo {author} {\bibfnamefont {N.}~\bibnamefont {Hansen}}\ and\ \bibinfo {author} {\bibfnamefont {W.~F.}\ \bibnamefont {Van~Gunsteren}},\ }\bibfield  {title} {\enquote {\bibinfo {title} {Practical aspects of free-energy calculations: a review},}\ }\href@noop {} {\bibfield  {journal} {\bibinfo  {journal} {J. Chem. Theory Comput.}\ }\textbf {\bibinfo {volume} {10}},\ \bibinfo {pages} {2632--2647} (\bibinfo {year} {2014})}\BibitemShut {NoStop}%
\bibitem [{\citenamefont {Frenkel}\ and\ \citenamefont {Smit}(2023)}]{frenkel2023understanding}%
  \BibitemOpen
  \bibfield  {author} {\bibinfo {author} {\bibfnamefont {D.}~\bibnamefont {Frenkel}}\ and\ \bibinfo {author} {\bibfnamefont {B.}~\bibnamefont {Smit}},\ }\href@noop {} {\emph {\bibinfo {title} {Understanding molecular simulation: from algorithms to applications}}}\ (\bibinfo  {publisher} {Elsevier},\ \bibinfo {year} {2023})\BibitemShut {NoStop}%
\bibitem [{\citenamefont {Best}\ and\ \citenamefont {Hummer}(2005)}]{best2005reaction}%
  \BibitemOpen
  \bibfield  {author} {\bibinfo {author} {\bibfnamefont {R.~B.}\ \bibnamefont {Best}}\ and\ \bibinfo {author} {\bibfnamefont {G.}~\bibnamefont {Hummer}},\ }\bibfield  {title} {\enquote {\bibinfo {title} {Reaction coordinates and rates from transition paths},}\ }\href {https://doi.org/10.1073/pnas.0408098102} {\bibfield  {journal} {\bibinfo  {journal} {Proc. Natl. Acad. Sci.}\ }\textbf {\bibinfo {volume} {102}},\ \bibinfo {pages} {6732–6737} (\bibinfo {year} {2005})}\BibitemShut {NoStop}%
\bibitem [{\citenamefont {Laio}\ and\ \citenamefont {Parrinello}(2002)}]{laio2002escaping}%
  \BibitemOpen
  \bibfield  {author} {\bibinfo {author} {\bibfnamefont {A.}~\bibnamefont {Laio}}\ and\ \bibinfo {author} {\bibfnamefont {M.}~\bibnamefont {Parrinello}},\ }\bibfield  {title} {\enquote {\bibinfo {title} {Escaping free-energy minima},}\ }\href@noop {} {\bibfield  {journal} {\bibinfo  {journal} {Proc. Natl. Acad. Sci.}\ }\textbf {\bibinfo {volume} {99}},\ \bibinfo {pages} {12562--12566} (\bibinfo {year} {2002})}\BibitemShut {NoStop}%
\bibitem [{\citenamefont {Darve}, \citenamefont {Rodr{\'\i}guez-G{\'o}mez},\ and\ \citenamefont {Pohorille}(2008)}]{darve2008adaptive}%
  \BibitemOpen
  \bibfield  {author} {\bibinfo {author} {\bibfnamefont {E.}~\bibnamefont {Darve}}, \bibinfo {author} {\bibfnamefont {D.}~\bibnamefont {Rodr{\'\i}guez-G{\'o}mez}},\ and\ \bibinfo {author} {\bibfnamefont {A.}~\bibnamefont {Pohorille}},\ }\bibfield  {title} {\enquote {\bibinfo {title} {Adaptive biasing force method for scalar and vector free energy calculations},}\ }\href@noop {} {\bibfield  {journal} {\bibinfo  {journal} {J. Chem. Phys.}\ }\textbf {\bibinfo {volume} {128}},\ \bibinfo {pages} {144120} (\bibinfo {year} {2008})}\BibitemShut {NoStop}%
\bibitem [{\citenamefont {Sugita}\ and\ \citenamefont {Okamoto}(1999)}]{sugita1999replica}%
  \BibitemOpen
  \bibfield  {author} {\bibinfo {author} {\bibfnamefont {Y.}~\bibnamefont {Sugita}}\ and\ \bibinfo {author} {\bibfnamefont {Y.}~\bibnamefont {Okamoto}},\ }\bibfield  {title} {\enquote {\bibinfo {title} {Replica-exchange molecular dynamics method for protein folding},}\ }\href@noop {} {\bibfield  {journal} {\bibinfo  {journal} {Chem. Phys. Lett.}\ }\textbf {\bibinfo {volume} {314}},\ \bibinfo {pages} {141--151} (\bibinfo {year} {1999})}\BibitemShut {NoStop}%
\bibitem [{\citenamefont {Hamelberg}, \citenamefont {Mongan},\ and\ \citenamefont {McCammon}(2004)}]{hamelberg2004accelerated}%
  \BibitemOpen
  \bibfield  {author} {\bibinfo {author} {\bibfnamefont {D.}~\bibnamefont {Hamelberg}}, \bibinfo {author} {\bibfnamefont {J.}~\bibnamefont {Mongan}},\ and\ \bibinfo {author} {\bibfnamefont {J.~A.}\ \bibnamefont {McCammon}},\ }\bibfield  {title} {\enquote {\bibinfo {title} {Accelerated molecular dynamics: a promising and efficient simulation method for biomolecules},}\ }\href@noop {} {\bibfield  {journal} {\bibinfo  {journal} {J. Chem. Phys.}\ }\textbf {\bibinfo {volume} {120}},\ \bibinfo {pages} {11919--11929} (\bibinfo {year} {2004})}\BibitemShut {NoStop}%
\bibitem [{\citenamefont {Torrie}\ and\ \citenamefont {Valleau}(1977)}]{torrie1977nonphysical}%
  \BibitemOpen
  \bibfield  {author} {\bibinfo {author} {\bibfnamefont {G.~M.}\ \bibnamefont {Torrie}}\ and\ \bibinfo {author} {\bibfnamefont {J.~P.}\ \bibnamefont {Valleau}},\ }\bibfield  {title} {\enquote {\bibinfo {title} {Nonphysical sampling distributions in monte carlo free-energy estimation: Umbrella sampling},}\ }\href@noop {} {\bibfield  {journal} {\bibinfo  {journal} {J. Comput. Phys.}\ }\textbf {\bibinfo {volume} {23}},\ \bibinfo {pages} {187--199} (\bibinfo {year} {1977})}\BibitemShut {NoStop}%
\bibitem [{\citenamefont {Roux}(1995)}]{roux1995calculation}%
  \BibitemOpen
  \bibfield  {author} {\bibinfo {author} {\bibfnamefont {B.}~\bibnamefont {Roux}},\ }\bibfield  {title} {\enquote {\bibinfo {title} {The calculation of the potential of mean force using computer simulations},}\ }\href@noop {} {\bibfield  {journal} {\bibinfo  {journal} {Comput. Phys. Commun.}\ }\textbf {\bibinfo {volume} {91}},\ \bibinfo {pages} {275--282} (\bibinfo {year} {1995})}\BibitemShut {NoStop}%
\bibitem [{\citenamefont {Kumar}\ \emph {et~al.}(1992)\citenamefont {Kumar}, \citenamefont {Rosenberg}, \citenamefont {Bouzida}, \citenamefont {Swendsen},\ and\ \citenamefont {Kollman}}]{kumar92}%
  \BibitemOpen
  \bibfield  {author} {\bibinfo {author} {\bibfnamefont {S.}~\bibnamefont {Kumar}}, \bibinfo {author} {\bibfnamefont {J.~M.}\ \bibnamefont {Rosenberg}}, \bibinfo {author} {\bibfnamefont {D.}~\bibnamefont {Bouzida}}, \bibinfo {author} {\bibfnamefont {R.~H.}\ \bibnamefont {Swendsen}},\ and\ \bibinfo {author} {\bibfnamefont {P.~A.}\ \bibnamefont {Kollman}},\ }\bibfield  {title} {\enquote {\bibinfo {title} {The weighted histogram analysis method for free--energy calculations on biomolecules. i. the method},}\ }\href {https://doi.org/10.1002/jcc.540130812} {\bibfield  {journal} {\bibinfo  {journal} {J. Comput. Chem.}\ }\textbf {\bibinfo {volume} {13}},\ \bibinfo {pages} {1011--1021} (\bibinfo {year} {1992})}\BibitemShut {NoStop}%
\bibitem [{\citenamefont {Hub}, \citenamefont {de~Groot},\ and\ \citenamefont {van~der Spoel}(2010)}]{hub10}%
  \BibitemOpen
  \bibfield  {author} {\bibinfo {author} {\bibfnamefont {J.~S.}\ \bibnamefont {Hub}}, \bibinfo {author} {\bibfnamefont {B.~L.}\ \bibnamefont {de~Groot}},\ and\ \bibinfo {author} {\bibfnamefont {D.}~\bibnamefont {van~der Spoel}},\ }\bibfield  {title} {\enquote {\bibinfo {title} {g\textunderscore{}wham---a free weighted histogram analysis implementation including robust error and autocorrelation estimates},}\ }\href {https://doi.org/10.1021/ct100494z} {\bibfield  {journal} {\bibinfo  {journal} {J. Chem. Theory Comput.}\ }\textbf {\bibinfo {volume} {6}},\ \bibinfo {pages} {3713--3720} (\bibinfo {year} {2010})}\BibitemShut {NoStop}%
\bibitem [{\citenamefont {Shirts}\ and\ \citenamefont {Chodera}(2008)}]{shirts2008}%
  \BibitemOpen
  \bibfield  {author} {\bibinfo {author} {\bibfnamefont {M.~R.}\ \bibnamefont {Shirts}}\ and\ \bibinfo {author} {\bibfnamefont {J.~D.}\ \bibnamefont {Chodera}},\ }\bibfield  {title} {\enquote {\bibinfo {title} {Statistically optimal analysis of samples from multiple equilibrium states},}\ }\href {https://doi.org/10.1063/1.2978177} {\bibfield  {journal} {\bibinfo  {journal} {J. Chem. Phys.}\ }\textbf {\bibinfo {volume} {129}},\ \bibinfo {pages} {124105} (\bibinfo {year} {2008})}\BibitemShut {NoStop}%
\bibitem [{\citenamefont {Orsi}\ and\ \citenamefont {Essex}(2010)}]{orsi2010passive}%
  \BibitemOpen
  \bibfield  {author} {\bibinfo {author} {\bibfnamefont {M.}~\bibnamefont {Orsi}}\ and\ \bibinfo {author} {\bibfnamefont {J.~W.}\ \bibnamefont {Essex}},\ }\bibfield  {title} {\enquote {\bibinfo {title} {Passive permeation across lipid bilayers: a literature review},}\ }\href@noop {} {\bibfield  {journal} {\bibinfo  {journal} {Mol. Simul. Biomembr. Biophys. To Funct.}\ ,\ \bibinfo {pages} {76}} (\bibinfo {year} {2010})}\BibitemShut {NoStop}%
\bibitem [{\citenamefont {Buch}, \citenamefont {Sadiq},\ and\ \citenamefont {De~Fabritiis}(2011)}]{buch2011optimized}%
  \BibitemOpen
  \bibfield  {author} {\bibinfo {author} {\bibfnamefont {I.}~\bibnamefont {Buch}}, \bibinfo {author} {\bibfnamefont {S.~K.}\ \bibnamefont {Sadiq}},\ and\ \bibinfo {author} {\bibfnamefont {G.}~\bibnamefont {De~Fabritiis}},\ }\bibfield  {title} {\enquote {\bibinfo {title} {Optimized potential of mean force calculations for standard binding free energies},}\ }\href@noop {} {\bibfield  {journal} {\bibinfo  {journal} {J. Chem. Theory Comput.}\ }\textbf {\bibinfo {volume} {7}},\ \bibinfo {pages} {1765--1772} (\bibinfo {year} {2011})}\BibitemShut {NoStop}%
\bibitem [{\citenamefont {Swift}\ and\ \citenamefont {Amaro}(2013)}]{swift2013back}%
  \BibitemOpen
  \bibfield  {author} {\bibinfo {author} {\bibfnamefont {R.~V.}\ \bibnamefont {Swift}}\ and\ \bibinfo {author} {\bibfnamefont {R.~E.}\ \bibnamefont {Amaro}},\ }\bibfield  {title} {\enquote {\bibinfo {title} {Back to the future: can physical models of passive membrane permeability help reduce drug candidate attrition and move us beyond qspr?}}\ }\href@noop {} {\bibfield  {journal} {\bibinfo  {journal} {Chem. Biol. \& Drug Des.}\ }\textbf {\bibinfo {volume} {81}},\ \bibinfo {pages} {61--71} (\bibinfo {year} {2013})}\BibitemShut {NoStop}%
\bibitem [{\citenamefont {Carpenter}\ \emph {et~al.}(2014)\citenamefont {Carpenter}, \citenamefont {Kirshner}, \citenamefont {Lau}, \citenamefont {Wong}, \citenamefont {Nilmeier},\ and\ \citenamefont {Lightstone}}]{carpenter2014method}%
  \BibitemOpen
  \bibfield  {author} {\bibinfo {author} {\bibfnamefont {T.~S.}\ \bibnamefont {Carpenter}}, \bibinfo {author} {\bibfnamefont {D.~A.}\ \bibnamefont {Kirshner}}, \bibinfo {author} {\bibfnamefont {E.~Y.}\ \bibnamefont {Lau}}, \bibinfo {author} {\bibfnamefont {S.~E.}\ \bibnamefont {Wong}}, \bibinfo {author} {\bibfnamefont {J.~P.}\ \bibnamefont {Nilmeier}},\ and\ \bibinfo {author} {\bibfnamefont {F.~C.}\ \bibnamefont {Lightstone}},\ }\bibfield  {title} {\enquote {\bibinfo {title} {A method to predict blood-brain barrier permeability of drug-like compounds using molecular dynamics simulations},}\ }\href@noop {} {\bibfield  {journal} {\bibinfo  {journal} {Biophys. J.}\ }\textbf {\bibinfo {volume} {107}},\ \bibinfo {pages} {630--641} (\bibinfo {year} {2014})}\BibitemShut {NoStop}%
\bibitem [{\citenamefont {Lee}\ \emph {et~al.}(2016)\citenamefont {Lee}, \citenamefont {Comer}, \citenamefont {Herndon}, \citenamefont {Leung}, \citenamefont {Pavlova}, \citenamefont {Swift}, \citenamefont {Tung}, \citenamefont {Rowley}, \citenamefont {Amaro}, \citenamefont {Chipot} \emph {et~al.}}]{lee2016simulation}%
  \BibitemOpen
  \bibfield  {author} {\bibinfo {author} {\bibfnamefont {C.~T.}\ \bibnamefont {Lee}}, \bibinfo {author} {\bibfnamefont {J.}~\bibnamefont {Comer}}, \bibinfo {author} {\bibfnamefont {C.}~\bibnamefont {Herndon}}, \bibinfo {author} {\bibfnamefont {N.}~\bibnamefont {Leung}}, \bibinfo {author} {\bibfnamefont {A.}~\bibnamefont {Pavlova}}, \bibinfo {author} {\bibfnamefont {R.~V.}\ \bibnamefont {Swift}}, \bibinfo {author} {\bibfnamefont {C.}~\bibnamefont {Tung}}, \bibinfo {author} {\bibfnamefont {C.~N.}\ \bibnamefont {Rowley}}, \bibinfo {author} {\bibfnamefont {R.~E.}\ \bibnamefont {Amaro}}, \bibinfo {author} {\bibfnamefont {C.}~\bibnamefont {Chipot}}, \emph {et~al.},\ }\bibfield  {title} {\enquote {\bibinfo {title} {Simulation-based approaches for determining membrane permeability of small compounds},}\ }\href@noop {} {\bibfield  {journal} {\bibinfo  {journal} {J. Chem. Inf. Model.}\ }\textbf {\bibinfo {volume} {56}},\ \bibinfo {pages} {721--733} (\bibinfo {year} {2016})}\BibitemShut {NoStop}%
\bibitem [{\citenamefont {Bennion}\ \emph {et~al.}(2017)\citenamefont {Bennion}, \citenamefont {Be}, \citenamefont {McNerney}, \citenamefont {Lao}, \citenamefont {Carlson}, \citenamefont {Valdez}, \citenamefont {Malfatti}, \citenamefont {Enright}, \citenamefont {Nguyen}, \citenamefont {Lightstone} \emph {et~al.}}]{bennion2017predicting}%
  \BibitemOpen
  \bibfield  {author} {\bibinfo {author} {\bibfnamefont {B.~J.}\ \bibnamefont {Bennion}}, \bibinfo {author} {\bibfnamefont {N.~A.}\ \bibnamefont {Be}}, \bibinfo {author} {\bibfnamefont {M.~W.}\ \bibnamefont {McNerney}}, \bibinfo {author} {\bibfnamefont {V.}~\bibnamefont {Lao}}, \bibinfo {author} {\bibfnamefont {E.~M.}\ \bibnamefont {Carlson}}, \bibinfo {author} {\bibfnamefont {C.~A.}\ \bibnamefont {Valdez}}, \bibinfo {author} {\bibfnamefont {M.~A.}\ \bibnamefont {Malfatti}}, \bibinfo {author} {\bibfnamefont {H.~A.}\ \bibnamefont {Enright}}, \bibinfo {author} {\bibfnamefont {T.~H.}\ \bibnamefont {Nguyen}}, \bibinfo {author} {\bibfnamefont {F.~C.}\ \bibnamefont {Lightstone}}, \emph {et~al.},\ }\bibfield  {title} {\enquote {\bibinfo {title} {Predicting a drug's membrane permeability: A computational model validated with in vitro permeability assay data},}\ }\href@noop {} {\bibfield  {journal} {\bibinfo  {journal} {J. Phys. Chem. B}\ }\textbf {\bibinfo {volume} {121}},\ \bibinfo {pages} {5228--5237} (\bibinfo {year} {2017})}\BibitemShut {NoStop}%
\bibitem [{\citenamefont {Tse}\ \emph {et~al.}(2018)\citenamefont {Tse}, \citenamefont {Comer}, \citenamefont {Wang},\ and\ \citenamefont {Chipot}}]{tse2018link}%
  \BibitemOpen
  \bibfield  {author} {\bibinfo {author} {\bibfnamefont {C.~H.}\ \bibnamefont {Tse}}, \bibinfo {author} {\bibfnamefont {J.}~\bibnamefont {Comer}}, \bibinfo {author} {\bibfnamefont {Y.}~\bibnamefont {Wang}},\ and\ \bibinfo {author} {\bibfnamefont {C.}~\bibnamefont {Chipot}},\ }\bibfield  {title} {\enquote {\bibinfo {title} {Link between membrane composition and permeability to drugs},}\ }\href@noop {} {\bibfield  {journal} {\bibinfo  {journal} {J. Chem. Theory Comput.}\ }\textbf {\bibinfo {volume} {14}},\ \bibinfo {pages} {2895--2909} (\bibinfo {year} {2018})}\BibitemShut {NoStop}%
\bibitem [{\citenamefont {Menichetti}\ \emph {et~al.}(2017)\citenamefont {Menichetti}, \citenamefont {Kanekal}, \citenamefont {Kremer},\ and\ \citenamefont {Bereau}}]{menichetti2017silico}%
  \BibitemOpen
  \bibfield  {author} {\bibinfo {author} {\bibfnamefont {R.}~\bibnamefont {Menichetti}}, \bibinfo {author} {\bibfnamefont {K.~H.}\ \bibnamefont {Kanekal}}, \bibinfo {author} {\bibfnamefont {K.}~\bibnamefont {Kremer}},\ and\ \bibinfo {author} {\bibfnamefont {T.}~\bibnamefont {Bereau}},\ }\bibfield  {title} {\enquote {\bibinfo {title} {In silico screening of drug-membrane thermodynamics reveals linear relations between bulk partitioning and the potential of mean force},}\ }\href@noop {} {\bibfield  {journal} {\bibinfo  {journal} {J. Chem. Phys.}\ }\textbf {\bibinfo {volume} {147}},\ \bibinfo {pages} {125101} (\bibinfo {year} {2017})}\BibitemShut {NoStop}%
\bibitem [{\citenamefont {Menichetti}, \citenamefont {Kanekal},\ and\ \citenamefont {Bereau}(2019)}]{menichetti2019drug}%
  \BibitemOpen
  \bibfield  {author} {\bibinfo {author} {\bibfnamefont {R.}~\bibnamefont {Menichetti}}, \bibinfo {author} {\bibfnamefont {K.~H.}\ \bibnamefont {Kanekal}},\ and\ \bibinfo {author} {\bibfnamefont {T.}~\bibnamefont {Bereau}},\ }\bibfield  {title} {\enquote {\bibinfo {title} {Drug--membrane permeability across chemical space},}\ }\href@noop {} {\bibfield  {journal} {\bibinfo  {journal} {ACS Cent. Sci.}\ }\textbf {\bibinfo {volume} {5}},\ \bibinfo {pages} {290--298} (\bibinfo {year} {2019})}\BibitemShut {NoStop}%
\bibitem [{\citenamefont {Jarzynski}(1997)}]{jarzynski1997nonequilibrium}%
  \BibitemOpen
  \bibfield  {author} {\bibinfo {author} {\bibfnamefont {C.}~\bibnamefont {Jarzynski}},\ }\bibfield  {title} {\enquote {\bibinfo {title} {Nonequilibrium equality for free energy differences},}\ }\href@noop {} {\bibfield  {journal} {\bibinfo  {journal} {Phys. Rev. Lett.}\ }\textbf {\bibinfo {volume} {78}},\ \bibinfo {pages} {2690} (\bibinfo {year} {1997})}\BibitemShut {NoStop}%
\bibitem [{\citenamefont {Nagel}\ and\ \citenamefont {Bereau}(2025)}]{nagel2025fokker}%
  \BibitemOpen
  \bibfield  {author} {\bibinfo {author} {\bibfnamefont {D.}~\bibnamefont {Nagel}}\ and\ \bibinfo {author} {\bibfnamefont {T.}~\bibnamefont {Bereau}},\ }\bibfield  {title} {\enquote {\bibinfo {title} {Fokker--planck score learning: Efficient free-energy estimation under periodic boundary conditions},}\ }\href {https://doi.org/10.1021/acs.jpcb.5c04579} {\bibfield  {journal} {\bibinfo  {journal} {J. Phys. Chem. B}\ }\textbf {\bibinfo {volume} {129}},\ \bibinfo {pages} {11780--11790} (\bibinfo {year} {2025})}\BibitemShut {NoStop}%
\bibitem [{\citenamefont {Song}\ \emph {et~al.}(2020)\citenamefont {Song}, \citenamefont {Sohl-Dickstein}, \citenamefont {Kingma}, \citenamefont {Kumar}, \citenamefont {Ermon},\ and\ \citenamefont {Poole}}]{song2020score}%
  \BibitemOpen
  \bibfield  {author} {\bibinfo {author} {\bibfnamefont {Y.}~\bibnamefont {Song}}, \bibinfo {author} {\bibfnamefont {J.}~\bibnamefont {Sohl-Dickstein}}, \bibinfo {author} {\bibfnamefont {D.~P.}\ \bibnamefont {Kingma}}, \bibinfo {author} {\bibfnamefont {A.}~\bibnamefont {Kumar}}, \bibinfo {author} {\bibfnamefont {S.}~\bibnamefont {Ermon}},\ and\ \bibinfo {author} {\bibfnamefont {B.}~\bibnamefont {Poole}},\ }\bibfield  {title} {\enquote {\bibinfo {title} {Score-based generative modeling through stochastic differential equations},}\ }\href {https://doi.org/10.48550/arXiv.2011.13456} {\bibfield  {journal} {\bibinfo  {journal} {arXiv cs.LG}\ } (\bibinfo {year} {2020}),\ 10.48550/arXiv.2011.13456}\BibitemShut {NoStop}%
\bibitem [{\citenamefont {Risken}(1996)}]{risken1996fokker}%
  \BibitemOpen
  \bibfield  {author} {\bibinfo {author} {\bibfnamefont {H.}~\bibnamefont {Risken}},\ }\href@noop {} {\emph {\bibinfo {title} {Fokker-planck equation}}}\ (\bibinfo  {publisher} {Springer},\ \bibinfo {year} {1996})\BibitemShut {NoStop}%
\bibitem [{\citenamefont {Sohl-Dickstein}\ \emph {et~al.}(2015)\citenamefont {Sohl-Dickstein}, \citenamefont {Weiss}, \citenamefont {Maheswaranathan},\ and\ \citenamefont {Ganguli}}]{sohldickstein15}%
  \BibitemOpen
  \bibfield  {author} {\bibinfo {author} {\bibfnamefont {J.}~\bibnamefont {Sohl-Dickstein}}, \bibinfo {author} {\bibfnamefont {E.~A.}\ \bibnamefont {Weiss}}, \bibinfo {author} {\bibfnamefont {N.}~\bibnamefont {Maheswaranathan}},\ and\ \bibinfo {author} {\bibfnamefont {S.}~\bibnamefont {Ganguli}},\ }\bibfield  {title} {\enquote {\bibinfo {title} {Deep unsupervised learning using nonequilibrium thermodynamics},}\ }\href {https://doi.org/10.48550/arXiv.1503.03585} {\bibfield  {journal} {\bibinfo  {journal} {arXiv cs.LG}\ } (\bibinfo {year} {2015}),\ 10.48550/arXiv.1503.03585}\BibitemShut {NoStop}%
\bibitem [{\citenamefont {Ho}, \citenamefont {Jain},\ and\ \citenamefont {Abbeel}(2020)}]{ho20}%
  \BibitemOpen
  \bibfield  {author} {\bibinfo {author} {\bibfnamefont {J.}~\bibnamefont {Ho}}, \bibinfo {author} {\bibfnamefont {A.}~\bibnamefont {Jain}},\ and\ \bibinfo {author} {\bibfnamefont {P.}~\bibnamefont {Abbeel}},\ }\bibfield  {title} {\enquote {\bibinfo {title} {Denoising diffusion probabilistic models},}\ }\href {https://doi.org/10.48550/arXiv.2006.11239} {\bibfield  {journal} {\bibinfo  {journal} {arXiv cs.LG}\ } (\bibinfo {year} {2020}),\ 10.48550/arXiv.2006.11239}\BibitemShut {NoStop}%
\bibitem [{\citenamefont {Anderson}(1982)}]{anderson82}%
  \BibitemOpen
  \bibfield  {author} {\bibinfo {author} {\bibfnamefont {B.~D.}\ \bibnamefont {Anderson}},\ }\bibfield  {title} {\enquote {\bibinfo {title} {Reverse-time diffusion equation models},}\ }\href {https://doi.org/10.1016/0304-4149(82)90051-5} {\bibfield  {journal} {\bibinfo  {journal} {Stoch. Process. Their Appl.}\ }\textbf {\bibinfo {volume} {12}},\ \bibinfo {pages} {313--326} (\bibinfo {year} {1982})}\BibitemShut {NoStop}%
\bibitem [{\citenamefont {Arts}\ \emph {et~al.}(2023)\citenamefont {Arts}, \citenamefont {Garcia~Satorras}, \citenamefont {Huang}, \citenamefont {Zugner}, \citenamefont {Federici}, \citenamefont {Clementi}, \citenamefont {No{\'e}}, \citenamefont {Pinsler},\ and\ \citenamefont {van~den Berg}}]{arts2023two}%
  \BibitemOpen
  \bibfield  {author} {\bibinfo {author} {\bibfnamefont {M.}~\bibnamefont {Arts}}, \bibinfo {author} {\bibfnamefont {V.}~\bibnamefont {Garcia~Satorras}}, \bibinfo {author} {\bibfnamefont {C.-W.}\ \bibnamefont {Huang}}, \bibinfo {author} {\bibfnamefont {D.}~\bibnamefont {Zugner}}, \bibinfo {author} {\bibfnamefont {M.}~\bibnamefont {Federici}}, \bibinfo {author} {\bibfnamefont {C.}~\bibnamefont {Clementi}}, \bibinfo {author} {\bibfnamefont {F.}~\bibnamefont {No{\'e}}}, \bibinfo {author} {\bibfnamefont {R.}~\bibnamefont {Pinsler}},\ and\ \bibinfo {author} {\bibfnamefont {R.}~\bibnamefont {van~den Berg}},\ }\bibfield  {title} {\enquote {\bibinfo {title} {Two for one: Diffusion models and force fields for coarse-grained molecular dynamics},}\ }\href@noop {} {\bibfield  {journal} {\bibinfo  {journal} {J. Chem. Theory Comput.}\ }\textbf {\bibinfo {volume} {19}},\ \bibinfo {pages} {6151--6159} (\bibinfo {year} {2023})}\BibitemShut {NoStop}%
\bibitem [{\citenamefont {Máté}, \citenamefont {Fleuret},\ and\ \citenamefont {Bereau}(2024)}]{mate24}%
  \BibitemOpen
  \bibfield  {author} {\bibinfo {author} {\bibfnamefont {B.}~\bibnamefont {Máté}}, \bibinfo {author} {\bibfnamefont {F.}~\bibnamefont {Fleuret}},\ and\ \bibinfo {author} {\bibfnamefont {T.}~\bibnamefont {Bereau}},\ }\bibfield  {title} {\enquote {\bibinfo {title} {Neural thermodynamic integration: Free energies from energy-based diffusion models},}\ }\href {https://doi.org/10.1021/acs.jpclett.4c01958} {\bibfield  {journal} {\bibinfo  {journal} {J. Phys. Chem. Lett.}\ }\textbf {\bibinfo {volume} {15}},\ \bibinfo {pages} {11395–11404} (\bibinfo {year} {2024})}\BibitemShut {NoStop}%
\bibitem [{\citenamefont {M{\'a}t{\'e}}, \citenamefont {Fleuret},\ and\ \citenamefont {Bereau}(2025)}]{mate2025solvation}%
  \BibitemOpen
  \bibfield  {author} {\bibinfo {author} {\bibfnamefont {B.}~\bibnamefont {M{\'a}t{\'e}}}, \bibinfo {author} {\bibfnamefont {F.}~\bibnamefont {Fleuret}},\ and\ \bibinfo {author} {\bibfnamefont {T.}~\bibnamefont {Bereau}},\ }\bibfield  {title} {\enquote {\bibinfo {title} {Solvation free energies from neural thermodynamic integration},}\ }\href@noop {} {\bibfield  {journal} {\bibinfo  {journal} {J. Chem. Phys.}\ }\textbf {\bibinfo {volume} {162}},\ \bibinfo {pages} {11395--11404} (\bibinfo {year} {2025})}\BibitemShut {NoStop}%
\bibitem [{\citenamefont {Corso}\ \emph {et~al.}(2022)\citenamefont {Corso}, \citenamefont {St\"{a}rk}, \citenamefont {Jing}, \citenamefont {Barzilay},\ and\ \citenamefont {Jaakkola}}]{corso2022diffdock}%
  \BibitemOpen
  \bibfield  {author} {\bibinfo {author} {\bibfnamefont {G.}~\bibnamefont {Corso}}, \bibinfo {author} {\bibfnamefont {H.}~\bibnamefont {St\"{a}rk}}, \bibinfo {author} {\bibfnamefont {B.}~\bibnamefont {Jing}}, \bibinfo {author} {\bibfnamefont {R.}~\bibnamefont {Barzilay}},\ and\ \bibinfo {author} {\bibfnamefont {T.}~\bibnamefont {Jaakkola}},\ }\bibfield  {title} {\enquote {\bibinfo {title} {Diffdock: Diffusion steps, twists, and turns for molecular docking},}\ }\href {https://doi.org/10.48550/arxiv.2210.01776} {\bibfield  {journal} {\bibinfo  {journal} {arXiv q-bio.BM}\ } (\bibinfo {year} {2022}),\ 10.48550/arxiv.2210.01776}\BibitemShut {NoStop}%
\bibitem [{\citenamefont {Yim}\ \emph {et~al.}(2023)\citenamefont {Yim}, \citenamefont {Trippe}, \citenamefont {De~Bortoli}, \citenamefont {Mathieu}, \citenamefont {Doucet}, \citenamefont {Barzilay},\ and\ \citenamefont {Jaakkola}}]{yim2023se3}%
  \BibitemOpen
  \bibfield  {author} {\bibinfo {author} {\bibfnamefont {J.}~\bibnamefont {Yim}}, \bibinfo {author} {\bibfnamefont {B.~L.}\ \bibnamefont {Trippe}}, \bibinfo {author} {\bibfnamefont {V.}~\bibnamefont {De~Bortoli}}, \bibinfo {author} {\bibfnamefont {E.}~\bibnamefont {Mathieu}}, \bibinfo {author} {\bibfnamefont {A.}~\bibnamefont {Doucet}}, \bibinfo {author} {\bibfnamefont {R.}~\bibnamefont {Barzilay}},\ and\ \bibinfo {author} {\bibfnamefont {T.}~\bibnamefont {Jaakkola}},\ }\bibfield  {title} {\enquote {\bibinfo {title} {Se(3) diffusion model with application to protein backbone generation},}\ }\href {https://doi.org/10.48550/arxiv.2302.02277} {\bibfield  {journal} {\bibinfo  {journal} {arXiv cs.LG}\ } (\bibinfo {year} {2023}),\ 10.48550/arxiv.2302.02277}\BibitemShut {NoStop}%
\bibitem [{\citenamefont {Plainer}\ \emph {et~al.}(2025)\citenamefont {Plainer}, \citenamefont {Wu}, \citenamefont {Klein}, \citenamefont {G\"{u}nnemann},\ and\ \citenamefont {Noé}}]{plainer2025consistent}%
  \BibitemOpen
  \bibfield  {author} {\bibinfo {author} {\bibfnamefont {M.}~\bibnamefont {Plainer}}, \bibinfo {author} {\bibfnamefont {H.}~\bibnamefont {Wu}}, \bibinfo {author} {\bibfnamefont {L.}~\bibnamefont {Klein}}, \bibinfo {author} {\bibfnamefont {S.}~\bibnamefont {G\"{u}nnemann}},\ and\ \bibinfo {author} {\bibfnamefont {F.}~\bibnamefont {Noé}},\ }\href {https://doi.org/10.48550/ARXIV.2506.17139} {\enquote {\bibinfo {title} {Consistent sampling and simulation: Molecular dynamics with energy-based diffusion models},}\ } (\bibinfo {year} {2025})\BibitemShut {NoStop}%
\bibitem [{\citenamefont {Souza}\ \emph {et~al.}(2021)\citenamefont {Souza}, \citenamefont {Alessandri}, \citenamefont {Barnoud}, \citenamefont {Thallmair}, \citenamefont {Faustino}, \citenamefont {Grünewald}, \citenamefont {Patmanidis}, \citenamefont {Abdizadeh}, \citenamefont {Bruininks}, \citenamefont {Wassenaar}, \citenamefont {Kroon}, \citenamefont {Melcr}, \citenamefont {Nieto}, \citenamefont {Corradi}, \citenamefont {Khan}, \citenamefont {Domański}, \citenamefont {Javanainen}, \citenamefont {Martinez-Seara}, \citenamefont {Reuter}, \citenamefont {Best}, \citenamefont {Vattulainen}, \citenamefont {Monticelli}, \citenamefont {Periole}, \citenamefont {Tieleman}, \citenamefont {de~Vries},\ and\ \citenamefont {Marrink}}]{souza21}%
  \BibitemOpen
  \bibfield  {author} {\bibinfo {author} {\bibfnamefont {P.~C.~T.}\ \bibnamefont {Souza}}, \bibinfo {author} {\bibfnamefont {R.}~\bibnamefont {Alessandri}}, \bibinfo {author} {\bibfnamefont {J.}~\bibnamefont {Barnoud}}, \bibinfo {author} {\bibfnamefont {S.}~\bibnamefont {Thallmair}}, \bibinfo {author} {\bibfnamefont {I.}~\bibnamefont {Faustino}}, \bibinfo {author} {\bibfnamefont {F.}~\bibnamefont {Grünewald}}, \bibinfo {author} {\bibfnamefont {I.}~\bibnamefont {Patmanidis}}, \bibinfo {author} {\bibfnamefont {H.}~\bibnamefont {Abdizadeh}}, \bibinfo {author} {\bibfnamefont {B.~M.~H.}\ \bibnamefont {Bruininks}}, \bibinfo {author} {\bibfnamefont {T.~A.}\ \bibnamefont {Wassenaar}}, \bibinfo {author} {\bibfnamefont {P.~C.}\ \bibnamefont {Kroon}}, \bibinfo {author} {\bibfnamefont {J.}~\bibnamefont {Melcr}}, \bibinfo {author} {\bibfnamefont {V.}~\bibnamefont {Nieto}}, \bibinfo {author} {\bibfnamefont {V.}~\bibnamefont {Corradi}}, \bibinfo {author} {\bibfnamefont {H.~M.}\ \bibnamefont {Khan}}, \bibinfo {author} {\bibfnamefont {J.}~\bibnamefont {Domański}}, \bibinfo {author} {\bibfnamefont {M.}~\bibnamefont {Javanainen}}, \bibinfo {author} {\bibfnamefont {H.}~\bibnamefont {Martinez-Seara}}, \bibinfo {author} {\bibfnamefont {N.}~\bibnamefont {Reuter}}, \bibinfo {author} {\bibfnamefont {R.~B.}\ \bibnamefont {Best}}, \bibinfo {author} {\bibfnamefont {I.}~\bibnamefont {Vattulainen}}, \bibinfo {author} {\bibfnamefont {L.}~\bibnamefont {Monticelli}}, \bibinfo {author} {\bibfnamefont {X.}~\bibnamefont {Periole}}, \bibinfo {author} {\bibfnamefont {D.~P.}\ \bibnamefont {Tieleman}}, \bibinfo {author} {\bibfnamefont {A.~H.}\ \bibnamefont {de~Vries}},\ and\ \bibinfo {author} {\bibfnamefont {S.~J.}\ \bibnamefont {Marrink}},\ }\bibfield  {title} {\enquote {\bibinfo {title} {Martini 3: a general purpose force field for coarse-grained molecular dynamics},}\ }\href {https://doi.org/10.1038/s41592-021-01098-3} {\bibfield  {journal} {\bibinfo  {journal} {Nat. Method}\ }\textbf {\bibinfo {volume} {18}},\ \bibinfo {pages} {382–388} (\bibinfo {year} {2021})}\BibitemShut {NoStop}%
\bibitem [{\citenamefont {Bereau}, \citenamefont {Walter},\ and\ \citenamefont {Rudzinski}(2024)}]{martignac24}%
  \BibitemOpen
  \bibfield  {author} {\bibinfo {author} {\bibfnamefont {T.}~\bibnamefont {Bereau}}, \bibinfo {author} {\bibfnamefont {L.~J.}\ \bibnamefont {Walter}},\ and\ \bibinfo {author} {\bibfnamefont {J.~F.}\ \bibnamefont {Rudzinski}},\ }\bibfield  {title} {\enquote {\bibinfo {title} {Martignac: Computational workflows for reproducible, traceable, and composable coarse-grained martini simulations},}\ }\href {https://doi.org/10.1021/acs.jcim.4c01754} {\bibfield  {journal} {\bibinfo  {journal} {J. Chem. Inf. Model.}\ }\textbf {\bibinfo {volume} {64}},\ \bibinfo {pages} {9413–9423} (\bibinfo {year} {2024})}\BibitemShut {NoStop}%
\bibitem [{\citenamefont {Berendsen}, \citenamefont {van~der Spoel},\ and\ \citenamefont {van Drunen}(1995)}]{berendsen1995gromacs}%
  \BibitemOpen
  \bibfield  {author} {\bibinfo {author} {\bibfnamefont {H.~J.}\ \bibnamefont {Berendsen}}, \bibinfo {author} {\bibfnamefont {D.}~\bibnamefont {van~der Spoel}},\ and\ \bibinfo {author} {\bibfnamefont {R.}~\bibnamefont {van Drunen}},\ }\bibfield  {title} {\enquote {\bibinfo {title} {Gromacs: A message-passing parallel molecular dynamics implementation},}\ }\href@noop {} {\bibfield  {journal} {\bibinfo  {journal} {Comput. Phys. Commun.}\ }\textbf {\bibinfo {volume} {91}},\ \bibinfo {pages} {43--56} (\bibinfo {year} {1995})}\BibitemShut {NoStop}%
\bibitem [{\citenamefont {Bussi}, \citenamefont {Donadio},\ and\ \citenamefont {Parrinello}(2007)}]{bussi07}%
  \BibitemOpen
  \bibfield  {author} {\bibinfo {author} {\bibfnamefont {G.}~\bibnamefont {Bussi}}, \bibinfo {author} {\bibfnamefont {D.}~\bibnamefont {Donadio}},\ and\ \bibinfo {author} {\bibfnamefont {M.}~\bibnamefont {Parrinello}},\ }\bibfield  {title} {\enquote {\bibinfo {title} {Canonical sampling through velocity rescaling},}\ }\href {https://doi.org/10.1063/1.2408420} {\bibfield  {journal} {\bibinfo  {journal} {J. Chem. Phys.}\ }\textbf {\bibinfo {volume} {126}},\ \bibinfo {pages} {014101} (\bibinfo {year} {2007})}\BibitemShut {NoStop}%
\bibitem [{\citenamefont {Bernetti}\ and\ \citenamefont {Bussi}(2020)}]{bernetti20}%
  \BibitemOpen
  \bibfield  {author} {\bibinfo {author} {\bibfnamefont {M.}~\bibnamefont {Bernetti}}\ and\ \bibinfo {author} {\bibfnamefont {G.}~\bibnamefont {Bussi}},\ }\bibfield  {title} {\enquote {\bibinfo {title} {Pressure control using stochastic cell rescaling},}\ }\href {https://doi.org/10.1063/5.0020514} {\bibfield  {journal} {\bibinfo  {journal} {J. Chem. Phys.}\ }\textbf {\bibinfo {volume} {153}},\ \bibinfo {pages} {114107} (\bibinfo {year} {2020})}\BibitemShut {NoStop}%
\bibitem [{\citenamefont {consortium}(2019)}]{plumed2019}%
  \BibitemOpen
  \bibfield  {author} {\bibinfo {author} {\bibfnamefont {T.~P.}\ \bibnamefont {consortium}},\ }\bibfield  {title} {\enquote {\bibinfo {title} {Promoting transparency and reproducibility in enhanced molecular simulations},}\ }\href {https://doi.org/10.1038/s41592-019-0506-8} {\bibfield  {journal} {\bibinfo  {journal} {Nature Methods}\ }\textbf {\bibinfo {volume} {16}},\ \bibinfo {pages} {670–673} (\bibinfo {year} {2019})}\BibitemShut {NoStop}%
\bibitem [{\citenamefont {Vanommeslaeghe}\ \emph {et~al.}(2009)\citenamefont {Vanommeslaeghe}, \citenamefont {Hatcher}, \citenamefont {Acharya}, \citenamefont {Kundu}, \citenamefont {Zhong}, \citenamefont {Shim}, \citenamefont {Darian}, \citenamefont {Guvench}, \citenamefont {Lopes}, \citenamefont {Vorobyov},\ and\ \citenamefont {Mackerell}}]{vanommeslaeghe2009CGenFF}%
  \BibitemOpen
  \bibfield  {author} {\bibinfo {author} {\bibfnamefont {K.}~\bibnamefont {Vanommeslaeghe}}, \bibinfo {author} {\bibfnamefont {E.}~\bibnamefont {Hatcher}}, \bibinfo {author} {\bibfnamefont {C.}~\bibnamefont {Acharya}}, \bibinfo {author} {\bibfnamefont {S.}~\bibnamefont {Kundu}}, \bibinfo {author} {\bibfnamefont {S.}~\bibnamefont {Zhong}}, \bibinfo {author} {\bibfnamefont {J.}~\bibnamefont {Shim}}, \bibinfo {author} {\bibfnamefont {E.}~\bibnamefont {Darian}}, \bibinfo {author} {\bibfnamefont {O.}~\bibnamefont {Guvench}}, \bibinfo {author} {\bibfnamefont {P.}~\bibnamefont {Lopes}}, \bibinfo {author} {\bibfnamefont {I.}~\bibnamefont {Vorobyov}},\ and\ \bibinfo {author} {\bibfnamefont {A.~D.}\ \bibnamefont {Mackerell}},\ }\bibfield  {title} {\enquote {\bibinfo {title} {Charmm general force field: A force field for drug‐like molecules compatible with the charmm all‐atom additive biological force fields},}\ }\href {https://doi.org/10.1002/jcc.21367} {\bibfield  {journal} {\bibinfo  {journal} {J. Comput. Chem.}\ }\textbf {\bibinfo {volume} {31}},\ \bibinfo {pages} {671–690} (\bibinfo {year} {2009})}\BibitemShut {NoStop}%
\bibitem [{\citenamefont {Klauda}\ \emph {et~al.}(2010)\citenamefont {Klauda}, \citenamefont {Venable}, \citenamefont {Freites}, \citenamefont {O’Connor}, \citenamefont {Tobias}, \citenamefont {Mondragon-Ramirez}, \citenamefont {Vorobyov}, \citenamefont {MacKerell},\ and\ \citenamefont {Pastor}}]{klauda2010charmm36lipids}%
  \BibitemOpen
  \bibfield  {author} {\bibinfo {author} {\bibfnamefont {J.~B.}\ \bibnamefont {Klauda}}, \bibinfo {author} {\bibfnamefont {R.~M.}\ \bibnamefont {Venable}}, \bibinfo {author} {\bibfnamefont {J.~A.}\ \bibnamefont {Freites}}, \bibinfo {author} {\bibfnamefont {J.~W.}\ \bibnamefont {O’Connor}}, \bibinfo {author} {\bibfnamefont {D.~J.}\ \bibnamefont {Tobias}}, \bibinfo {author} {\bibfnamefont {C.}~\bibnamefont {Mondragon-Ramirez}}, \bibinfo {author} {\bibfnamefont {I.}~\bibnamefont {Vorobyov}}, \bibinfo {author} {\bibfnamefont {A.~D.}\ \bibnamefont {MacKerell}},\ and\ \bibinfo {author} {\bibfnamefont {R.~W.}\ \bibnamefont {Pastor}},\ }\bibfield  {title} {\enquote {\bibinfo {title} {Update of the charmm all-atom additive force field for lipids: Validation on six lipid types},}\ }\href {https://doi.org/10.1021/jp101759q} {\bibfield  {journal} {\bibinfo  {journal} {J. Phys. Chem. B}\ }\textbf {\bibinfo {volume} {114}},\ \bibinfo {pages} {7830–7843} (\bibinfo {year} {2010})}\BibitemShut {NoStop}%
\bibitem [{\citenamefont {Jo}\ \emph {et~al.}(2008)\citenamefont {Jo}, \citenamefont {Kim}, \citenamefont {Iyer},\ and\ \citenamefont {Im}}]{jo2008charmmgui}%
  \BibitemOpen
  \bibfield  {author} {\bibinfo {author} {\bibfnamefont {S.}~\bibnamefont {Jo}}, \bibinfo {author} {\bibfnamefont {T.}~\bibnamefont {Kim}}, \bibinfo {author} {\bibfnamefont {V.~G.}\ \bibnamefont {Iyer}},\ and\ \bibinfo {author} {\bibfnamefont {W.}~\bibnamefont {Im}},\ }\bibfield  {title} {\enquote {\bibinfo {title} {Charmm-gui: A web-based graphical user interface for charmm},}\ }\href {https://doi.org/10.1002/jcc.20945} {\bibfield  {journal} {\bibinfo  {journal} {J. Comput. Chem.}\ }\textbf {\bibinfo {volume} {29}},\ \bibinfo {pages} {1859–1865} (\bibinfo {year} {2008})}\BibitemShut {NoStop}%
\bibitem [{\citenamefont {Feng}\ \emph {et~al.}(2023)\citenamefont {Feng}, \citenamefont {Park}, \citenamefont {Choi},\ and\ \citenamefont {Im}}]{feng2023charmmgui}%
  \BibitemOpen
  \bibfield  {author} {\bibinfo {author} {\bibfnamefont {S.}~\bibnamefont {Feng}}, \bibinfo {author} {\bibfnamefont {S.}~\bibnamefont {Park}}, \bibinfo {author} {\bibfnamefont {Y.~K.}\ \bibnamefont {Choi}},\ and\ \bibinfo {author} {\bibfnamefont {W.}~\bibnamefont {Im}},\ }\bibfield  {title} {\enquote {\bibinfo {title} {Charmm-gui membrane builder: Past, current, and future developments and applications},}\ }\href {https://doi.org/10.1021/acs.jctc.2c01246} {\bibfield  {journal} {\bibinfo  {journal} {J. Chem. Theory Comput.}\ }\textbf {\bibinfo {volume} {19}},\ \bibinfo {pages} {2161–2185} (\bibinfo {year} {2023})}\BibitemShut {NoStop}%
\bibitem [{\citenamefont {Lee}\ \emph {et~al.}(2015)\citenamefont {Lee}, \citenamefont {Cheng}, \citenamefont {Swails}, \citenamefont {Yeom}, \citenamefont {Eastman}, \citenamefont {Lemkul}, \citenamefont {Wei}, \citenamefont {Buckner}, \citenamefont {Jeong}, \citenamefont {Qi}, \citenamefont {Jo}, \citenamefont {Pande}, \citenamefont {Case}, \citenamefont {Brooks}, \citenamefont {MacKerell}, \citenamefont {Klauda},\ and\ \citenamefont {Im}}]{lee2015charmmgui}%
  \BibitemOpen
  \bibfield  {author} {\bibinfo {author} {\bibfnamefont {J.}~\bibnamefont {Lee}}, \bibinfo {author} {\bibfnamefont {X.}~\bibnamefont {Cheng}}, \bibinfo {author} {\bibfnamefont {J.~M.}\ \bibnamefont {Swails}}, \bibinfo {author} {\bibfnamefont {M.~S.}\ \bibnamefont {Yeom}}, \bibinfo {author} {\bibfnamefont {P.~K.}\ \bibnamefont {Eastman}}, \bibinfo {author} {\bibfnamefont {J.~A.}\ \bibnamefont {Lemkul}}, \bibinfo {author} {\bibfnamefont {S.}~\bibnamefont {Wei}}, \bibinfo {author} {\bibfnamefont {J.}~\bibnamefont {Buckner}}, \bibinfo {author} {\bibfnamefont {J.~C.}\ \bibnamefont {Jeong}}, \bibinfo {author} {\bibfnamefont {Y.}~\bibnamefont {Qi}}, \bibinfo {author} {\bibfnamefont {S.}~\bibnamefont {Jo}}, \bibinfo {author} {\bibfnamefont {V.~S.}\ \bibnamefont {Pande}}, \bibinfo {author} {\bibfnamefont {D.~A.}\ \bibnamefont {Case}}, \bibinfo {author} {\bibfnamefont {C.~L.}\ \bibnamefont {Brooks}}, \bibinfo {author} {\bibfnamefont {A.~D.}\ \bibnamefont {MacKerell}}, \bibinfo {author} {\bibfnamefont {J.~B.}\ \bibnamefont {Klauda}},\ and\ \bibinfo {author} {\bibfnamefont {W.}~\bibnamefont {Im}},\ }\bibfield  {title} {\enquote {\bibinfo {title} {Charmm-gui input generator for namd, gromacs, amber, openmm, and charmm/openmm simulations using the charmm36 additive force field},}\ }\href {https://doi.org/10.1021/acs.jctc.5b00935} {\bibfield  {journal} {\bibinfo  {journal} {J. Chem. Theory Comput.}\ }\textbf {\bibinfo {volume} {12}},\ \bibinfo {pages} {405–413} (\bibinfo {year} {2015})}\BibitemShut {NoStop}%
\bibitem [{\citenamefont {Lindorff‐Larsen}\ \emph {et~al.}(2010)\citenamefont {Lindorff‐Larsen}, \citenamefont {Piana}, \citenamefont {Palmo}, \citenamefont {Maragakis}, \citenamefont {Klepeis}, \citenamefont {Dror},\ and\ \citenamefont {Shaw}}]{lindorfflarsen2010amberff99sbildn}%
  \BibitemOpen
  \bibfield  {author} {\bibinfo {author} {\bibfnamefont {K.}~\bibnamefont {Lindorff‐Larsen}}, \bibinfo {author} {\bibfnamefont {S.}~\bibnamefont {Piana}}, \bibinfo {author} {\bibfnamefont {K.}~\bibnamefont {Palmo}}, \bibinfo {author} {\bibfnamefont {P.}~\bibnamefont {Maragakis}}, \bibinfo {author} {\bibfnamefont {J.~L.}\ \bibnamefont {Klepeis}}, \bibinfo {author} {\bibfnamefont {R.~O.}\ \bibnamefont {Dror}},\ and\ \bibinfo {author} {\bibfnamefont {D.~E.}\ \bibnamefont {Shaw}},\ }\bibfield  {title} {\enquote {\bibinfo {title} {Improved side‐chain torsion potentials for the amber ff99sb protein force field},}\ }\href {https://doi.org/10.1002/prot.22711} {\bibfield  {journal} {\bibinfo  {journal} {Proteins: Structure, Function, and Bioinformatics}\ }\textbf {\bibinfo {volume} {78}},\ \bibinfo {pages} {1950–1958} (\bibinfo {year} {2010})}\BibitemShut {NoStop}%
\bibitem [{\citenamefont {Bradbury}\ \emph {et~al.}(2018)\citenamefont {Bradbury}, \citenamefont {Frostig}, \citenamefont {Hawkins}, \citenamefont {Johnson}, \citenamefont {Leary}, \citenamefont {Maclaurin}, \citenamefont {Necula}, \citenamefont {Paszke}, \citenamefont {Vander{P}las}, \citenamefont {Wanderman-{M}ilne},\ and\ \citenamefont {Zhang}}]{jax}%
  \BibitemOpen
  \bibfield  {author} {\bibinfo {author} {\bibfnamefont {J.}~\bibnamefont {Bradbury}}, \bibinfo {author} {\bibfnamefont {R.}~\bibnamefont {Frostig}}, \bibinfo {author} {\bibfnamefont {P.}~\bibnamefont {Hawkins}}, \bibinfo {author} {\bibfnamefont {M.~J.}\ \bibnamefont {Johnson}}, \bibinfo {author} {\bibfnamefont {C.}~\bibnamefont {Leary}}, \bibinfo {author} {\bibfnamefont {D.}~\bibnamefont {Maclaurin}}, \bibinfo {author} {\bibfnamefont {G.}~\bibnamefont {Necula}}, \bibinfo {author} {\bibfnamefont {A.}~\bibnamefont {Paszke}}, \bibinfo {author} {\bibfnamefont {J.}~\bibnamefont {Vander{P}las}}, \bibinfo {author} {\bibfnamefont {S.}~\bibnamefont {Wanderman-{M}ilne}},\ and\ \bibinfo {author} {\bibfnamefont {Q.}~\bibnamefont {Zhang}},\ }\href {http://github.com/jax-ml/jax} {\enquote {\bibinfo {title} {{JAX}: composable transformations of {P}ython+{N}um{P}y programs},}\ } (\bibinfo {year} {2018})\BibitemShut {NoStop}%
\bibitem [{\citenamefont {Heek}\ \emph {et~al.}(2024)\citenamefont {Heek}, \citenamefont {Levskaya}, \citenamefont {Oliver}, \citenamefont {Ritter}, \citenamefont {Rondepierre}, \citenamefont {Steiner},\ and\ \citenamefont {van {Z}ee}}]{flax}%
  \BibitemOpen
  \bibfield  {author} {\bibinfo {author} {\bibfnamefont {J.}~\bibnamefont {Heek}}, \bibinfo {author} {\bibfnamefont {A.}~\bibnamefont {Levskaya}}, \bibinfo {author} {\bibfnamefont {A.}~\bibnamefont {Oliver}}, \bibinfo {author} {\bibfnamefont {M.}~\bibnamefont {Ritter}}, \bibinfo {author} {\bibfnamefont {B.}~\bibnamefont {Rondepierre}}, \bibinfo {author} {\bibfnamefont {A.}~\bibnamefont {Steiner}},\ and\ \bibinfo {author} {\bibfnamefont {M.}~\bibnamefont {van {Z}ee}},\ }\href {http://github.com/google/flax} {\enquote {\bibinfo {title} {{F}lax: A neural network library and ecosystem for {JAX}},}\ } (\bibinfo {year} {2024}),\ \bibinfo {note} {accessed: 2025-09-26}\BibitemShut {NoStop}%
\bibitem [{\citenamefont {DeepMind}\ \emph {et~al.}(2020)\citenamefont {DeepMind}, \citenamefont {Babuschkin}, \citenamefont {Baumli}, \citenamefont {Bell}, \citenamefont {Bhupatiraju}, \citenamefont {Bruce}, \citenamefont {Buchlovsky}, \citenamefont {Budden}, \citenamefont {Cai}, \citenamefont {Clark}, \citenamefont {Danihelka}, \citenamefont {Dedieu}, \citenamefont {Fantacci}, \citenamefont {Godwin}, \citenamefont {Jones}, \citenamefont {Hemsley}, \citenamefont {Hennigan}, \citenamefont {Hessel}, \citenamefont {Hou}, \citenamefont {Kapturowski}, \citenamefont {Keck}, \citenamefont {Kemaev}, \citenamefont {King}, \citenamefont {Kunesch}, \citenamefont {Martens}, \citenamefont {Merzic}, \citenamefont {Mikulik}, \citenamefont {Norman}, \citenamefont {Papamakarios}, \citenamefont {Quan}, \citenamefont {Ring}, \citenamefont {Ruiz}, \citenamefont {Sanchez}, \citenamefont {Sartran}, \citenamefont {Schneider}, \citenamefont {Sezener}, \citenamefont {Spencer}, \citenamefont {Srinivasan}, \citenamefont {Stanojevi\'{c}}, \citenamefont {Stokowiec}, \citenamefont {Wang}, \citenamefont {Zhou},\ and\ \citenamefont {Viola}}]{optax}%
  \BibitemOpen
  \bibfield  {author} {\bibinfo {author} {\bibnamefont {DeepMind}}, \bibinfo {author} {\bibfnamefont {I.}~\bibnamefont {Babuschkin}}, \bibinfo {author} {\bibfnamefont {K.}~\bibnamefont {Baumli}}, \bibinfo {author} {\bibfnamefont {A.}~\bibnamefont {Bell}}, \bibinfo {author} {\bibfnamefont {S.}~\bibnamefont {Bhupatiraju}}, \bibinfo {author} {\bibfnamefont {J.}~\bibnamefont {Bruce}}, \bibinfo {author} {\bibfnamefont {P.}~\bibnamefont {Buchlovsky}}, \bibinfo {author} {\bibfnamefont {D.}~\bibnamefont {Budden}}, \bibinfo {author} {\bibfnamefont {T.}~\bibnamefont {Cai}}, \bibinfo {author} {\bibfnamefont {A.}~\bibnamefont {Clark}}, \bibinfo {author} {\bibfnamefont {I.}~\bibnamefont {Danihelka}}, \bibinfo {author} {\bibfnamefont {A.}~\bibnamefont {Dedieu}}, \bibinfo {author} {\bibfnamefont {C.}~\bibnamefont {Fantacci}}, \bibinfo {author} {\bibfnamefont {J.}~\bibnamefont {Godwin}}, \bibinfo {author} {\bibfnamefont {C.}~\bibnamefont {Jones}}, \bibinfo {author} {\bibfnamefont {R.}~\bibnamefont {Hemsley}}, \bibinfo {author} {\bibfnamefont {T.}~\bibnamefont {Hennigan}}, \bibinfo {author} {\bibfnamefont {M.}~\bibnamefont {Hessel}}, \bibinfo {author} {\bibfnamefont {S.}~\bibnamefont {Hou}}, \bibinfo {author} {\bibfnamefont {S.}~\bibnamefont {Kapturowski}}, \bibinfo {author} {\bibfnamefont {T.}~\bibnamefont {Keck}}, \bibinfo {author} {\bibfnamefont {I.}~\bibnamefont {Kemaev}}, \bibinfo {author} {\bibfnamefont {M.}~\bibnamefont {King}}, \bibinfo {author} {\bibfnamefont {M.}~\bibnamefont {Kunesch}}, \bibinfo {author} {\bibfnamefont {L.}~\bibnamefont {Martens}}, \bibinfo {author} {\bibfnamefont {H.}~\bibnamefont {Merzic}}, \bibinfo {author} {\bibfnamefont {V.}~\bibnamefont {Mikulik}}, \bibinfo {author} {\bibfnamefont {T.}~\bibnamefont {Norman}}, \bibinfo {author} {\bibfnamefont {G.}~\bibnamefont {Papamakarios}}, \bibinfo {author} {\bibfnamefont {J.}~\bibnamefont {Quan}}, \bibinfo {author} {\bibfnamefont {R.}~\bibnamefont {Ring}}, \bibinfo {author} {\bibfnamefont {F.}~\bibnamefont {Ruiz}}, \bibinfo {author} {\bibfnamefont {A.}~\bibnamefont {Sanchez}}, \bibinfo {author} {\bibfnamefont {L.}~\bibnamefont {Sartran}}, \bibinfo {author} {\bibfnamefont {R.}~\bibnamefont {Schneider}}, \bibinfo {author} {\bibfnamefont {E.}~\bibnamefont {Sezener}}, \bibinfo {author} {\bibfnamefont {S.}~\bibnamefont {Spencer}}, \bibinfo {author} {\bibfnamefont {S.}~\bibnamefont {Srinivasan}}, \bibinfo {author} {\bibfnamefont {M.}~\bibnamefont {Stanojevi\'{c}}}, \bibinfo {author} {\bibfnamefont {W.}~\bibnamefont {Stokowiec}}, \bibinfo {author} {\bibfnamefont {L.}~\bibnamefont {Wang}}, \bibinfo {author} {\bibfnamefont {G.}~\bibnamefont {Zhou}},\ and\ \bibinfo {author} {\bibfnamefont {F.}~\bibnamefont {Viola}},\ }\href {http://github.com/google-deepmind} {\enquote {\bibinfo {title} {The {D}eep{M}ind {JAX} {E}cosystem},}\ } (\bibinfo {year} {2020}),\ \bibinfo {note} {accessed: 2025-09-26}\BibitemShut {NoStop}%
\bibitem [{\citenamefont {Loshchilov}\ and\ \citenamefont {Hutter}(2017)}]{loshchilov17}%
  \BibitemOpen
  \bibfield  {author} {\bibinfo {author} {\bibfnamefont {I.}~\bibnamefont {Loshchilov}}\ and\ \bibinfo {author} {\bibfnamefont {F.}~\bibnamefont {Hutter}},\ }\bibfield  {title} {\enquote {\bibinfo {title} {Decoupled weight decay regularization},}\ }\href {https://doi.org/10.48550/arXiv.1711.05101} {\bibfield  {journal} {\bibinfo  {journal} {arXiv cs.LG}\ } (\bibinfo {year} {2017}),\ 10.48550/arXiv.1711.05101}\BibitemShut {NoStop}%
\bibitem [{\citenamefont {Ramachandran}, \citenamefont {Zoph},\ and\ \citenamefont {Le}(2017)}]{ramachandran17}%
  \BibitemOpen
  \bibfield  {author} {\bibinfo {author} {\bibfnamefont {P.}~\bibnamefont {Ramachandran}}, \bibinfo {author} {\bibfnamefont {B.}~\bibnamefont {Zoph}},\ and\ \bibinfo {author} {\bibfnamefont {Q.~V.}\ \bibnamefont {Le}},\ }\bibfield  {title} {\enquote {\bibinfo {title} {Searching for activation functions},}\ }\href {https://doi.org/10.48550/arXiv.1710.05941} {\bibfield  {journal} {\bibinfo  {journal} {arXiv cs.NE}\ } (\bibinfo {year} {2017}),\ 10.48550/arXiv.1710.05941}\BibitemShut {NoStop}%
\bibitem [{\citenamefont {Comer}, \citenamefont {Schulten},\ and\ \citenamefont {Chipot}(2014)}]{comer2014diffusive}%
  \BibitemOpen
  \bibfield  {author} {\bibinfo {author} {\bibfnamefont {J.}~\bibnamefont {Comer}}, \bibinfo {author} {\bibfnamefont {K.}~\bibnamefont {Schulten}},\ and\ \bibinfo {author} {\bibfnamefont {C.}~\bibnamefont {Chipot}},\ }\bibfield  {title} {\enquote {\bibinfo {title} {Diffusive models of membrane permeation with explicit orientational freedom},}\ }\href {https://doi.org/10.1021/ct500209j} {\bibfield  {journal} {\bibinfo  {journal} {J. Chem. Theory Comput.}\ }\textbf {\bibinfo {volume} {10}},\ \bibinfo {pages} {2710--2718} (\bibinfo {year} {2014})}\BibitemShut {NoStop}%
\bibitem [{\citenamefont {Ghorbani}\ \emph {et~al.}(2020)\citenamefont {Ghorbani}, \citenamefont {Wang}, \citenamefont {Kr\"{a}mer},\ and\ \citenamefont {Klauda}}]{ghorbani2020molecular}%
  \BibitemOpen
  \bibfield  {author} {\bibinfo {author} {\bibfnamefont {M.}~\bibnamefont {Ghorbani}}, \bibinfo {author} {\bibfnamefont {E.}~\bibnamefont {Wang}}, \bibinfo {author} {\bibfnamefont {A.}~\bibnamefont {Kr\"{a}mer}},\ and\ \bibinfo {author} {\bibfnamefont {J.~B.}\ \bibnamefont {Klauda}},\ }\bibfield  {title} {\enquote {\bibinfo {title} {Molecular dynamics simulations of ethanol permeation through single and double-lipid bilayers},}\ }\href {https://doi.org/10.1063/5.0013430} {\bibfield  {journal} {\bibinfo  {journal} {J. Chem. Phys.}\ }\textbf {\bibinfo {volume} {153}} (\bibinfo {year} {2020}),\ 10.1063/5.0013430}\BibitemShut {NoStop}%
\bibitem [{\citenamefont {Tse}\ \emph {et~al.}(2019)\citenamefont {Tse}, \citenamefont {Comer}, \citenamefont {Sang~Chu}, \citenamefont {Wang},\ and\ \citenamefont {Chipot}}]{tse2019affordable}%
  \BibitemOpen
  \bibfield  {author} {\bibinfo {author} {\bibfnamefont {C.~H.}\ \bibnamefont {Tse}}, \bibinfo {author} {\bibfnamefont {J.}~\bibnamefont {Comer}}, \bibinfo {author} {\bibfnamefont {S.~K.}\ \bibnamefont {Sang~Chu}}, \bibinfo {author} {\bibfnamefont {Y.}~\bibnamefont {Wang}},\ and\ \bibinfo {author} {\bibfnamefont {C.}~\bibnamefont {Chipot}},\ }\bibfield  {title} {\enquote {\bibinfo {title} {Affordable membrane permeability calculations: Permeation of short-chain alcohols through pure-lipid bilayers and a mammalian cell membrane},}\ }\href {https://doi.org/10.1021/acs.jctc.9b00022} {\bibfield  {journal} {\bibinfo  {journal} {J. Chem. Theory Comput.}\ }\textbf {\bibinfo {volume} {15}},\ \bibinfo {pages} {2913–2924} (\bibinfo {year} {2019})}\BibitemShut {NoStop}%
\bibitem [{\citenamefont {Kr\"{a}mer}\ \emph {et~al.}(2020)\citenamefont {Kr\"{a}mer}, \citenamefont {Ghysels}, \citenamefont {Wang}, \citenamefont {Venable}, \citenamefont {Klauda}, \citenamefont {Brooks},\ and\ \citenamefont {Pastor}}]{Kraemer2020membrane}%
  \BibitemOpen
  \bibfield  {author} {\bibinfo {author} {\bibfnamefont {A.}~\bibnamefont {Kr\"{a}mer}}, \bibinfo {author} {\bibfnamefont {A.}~\bibnamefont {Ghysels}}, \bibinfo {author} {\bibfnamefont {E.}~\bibnamefont {Wang}}, \bibinfo {author} {\bibfnamefont {R.~M.}\ \bibnamefont {Venable}}, \bibinfo {author} {\bibfnamefont {J.~B.}\ \bibnamefont {Klauda}}, \bibinfo {author} {\bibfnamefont {B.~R.}\ \bibnamefont {Brooks}},\ and\ \bibinfo {author} {\bibfnamefont {R.~W.}\ \bibnamefont {Pastor}},\ }\bibfield  {title} {\enquote {\bibinfo {title} {Membrane permeability of small molecules from unbiased molecular dynamics simulations},}\ }\href {https://doi.org/10.1063/5.0013429} {\bibfield  {journal} {\bibinfo  {journal} {J. Chem. Phys.}\ }\textbf {\bibinfo {volume} {153}} (\bibinfo {year} {2020}),\ 10.1063/5.0013429}\BibitemShut {NoStop}%
\end{thebibliography}%
